\documentclass[twocolumn,
 amsmath,amssymb,
 aps, physrev, floatfix
]{revtex4-2}

\usepackage{graphicx}
\usepackage{dcolumn}
\usepackage{bm}
\usepackage{definitions}  
\usepackage{xspace}
\usepackage{float} 
\usepackage{subcaption} 
\usepackage{hyperref}
\usepackage{placeins} 

\newcommand{\latfigheight}{0.45\textheight}

\begin{document}

\preprint{preprint for Phys. Rev. D}

\title{\textbf{Modified hadronic interactions in 3-dimensional simulations} 
}%

\author{Jiří Blažek}
\author{Jan Ebr}
 \email{Contact author: ebr@fzu.cz}
  \author{Jakub Vícha}
   \author{Eva dos Santos}
\affiliation{%
 Institute of Physics of the Czech Academy of Sciences, Prague, Czech Republic
}%

 \author{Tanguy Pierog}
 \author{Ralf Ulrich}
\affiliation{%
 Karlsruhe Institute of Technology, Institut für Astroteilchenphysik, Karlsruhe, Germany
}%

\begin{abstract}
We present a method to test the impact of ad-hoc modifications of some of the generic parameters of hadronic interactions -- cross section, elasticity, and multiplicity -- on any observable quantity using full 3-dimensional simulations of extensive air showers induced by ultra-high-energy cosmic rays. Our approach not only extends the existing 1-dimensional tools to three dimensions, but also introduces more flexible features to better respond to the needs of both theory and experiment. We first thoroughly validate the \conexD framework for the simulation of both longitudinal and lateral features of air showers, in particular for a non-standard configuration of the framework in which different energy thresholds for modifications are applied. Moreover, we show that the implementations of the ad-hoc modifications in this configuration are consistent with the previous one-dimensional simulations. Lastly, we discuss the importance of studying the interaction modifications in three dimensions and the effects of parallel modifications of multiple parameters.
\end{abstract}


\date{\today}

\maketitle

\section{Introduction} \label{introduction}

Cosmic rays of the highest energies, when hitting target nuclei in the uppermost layers of the atmosphere, offer a unique window into hadronic physics otherwise unreachable by the Earth--bound accelerators. The model description of the physical processes responsible for the development of the extensive air showers of the highest energies has long been plagued by discrepancies. Several experiments report a measured excess of muons at ground level, which increases with energy over several orders of magnitude  \cite{Soldin:2021wyv, ArteagaVelazquez:20236d}. Recently, data from the Pierre Auger Observatory indicate that the discrepancies are not only in the predicted number of muons at the ground, but also in the predicted depths of the shower maxima \cite{kuba}. This might suggest a much heavier mass composition of cosmic rays than the one deduced using unmodified models of hadronic interactions; thus improving the global consistency of observed data \cite{HeavyMetal}. 

We have previously investigated the modifications to the properties of hadronic interactions which are necessary to account for these discrepancies by simulating hadronic interactions with modified characteristics \cite{icrc2021,uhecr2022,icrc2023,rumunsko}. In this work, we establish solid foundations for the use of such simulations and show that the implementation of the modifications into a fully 3-dimensional framework is key to the proper interpretation of the data. Furthermore, we show that the parallel modification of different properties is not equivalent to a simple composition of individual changes in the predictions for observable quantities. 

In Section II, we define the treatment of ad hoc modifications of hadronic interactions, which is validated in Section III. We demonstrate the importance of the 3D simulation approach in Section IV and discuss the parallelism and commutativity of these modifications in Section V. Section VI then contains the conclusion of the results. 

\section{Modified Characteristics of Hadronic Interactions}

In \cite{Ulrich:2010rg}, the effects of various modifications of the properties of hadronic interactions on the development of air showers were studied using the \conex \cite{Bergmann:2006yz} simulation framework. Modifications of cross section, elasticity (i.e., the fraction of total energy carried by the most-energetic particle), multiplicity (i.e., the number of particles in an interaction), or pion charge ratio were applied to all hadronic interactions above a laboratory threshold energy $E_{\mathrm{thr}}$, with the magnitude of the modifications $f(E)$ increasing logarithmically with energy to reach a chosen value of $f_{19}$ at 10 EeV, according to:

\begin{equation}
    f(E, f_{19}) = 1 + (f_{19} - 1)\cdot F(E)
    \label{eq_f19}
\end{equation}
where $F(E) = 0$ below $E_{\mathrm{thr}}$ and otherwise
\begin{equation}
    F(E) = \frac{\mathrm{log}_{10}(E / E_{\mathrm{thr}})}{\mathrm{log}_{10}(\mathrm{10~EeV} / E_{\mathrm{thr}})}.
\end{equation}

The main limitation of this study stems from the 1-dimensional character of \conex, which provides no information about the 3-dimensional structure of the shower. This is particularly important for the comparison with any sparse arrays of detectors on the ground, which typically provide the measurement of particle numbers at some given distance from the shower axis, the behavior of which under the modifications may substantially differ from the behavior of the total number of particles available from \conex.

To study the effects of modifications of hadronic interactions on 3-dimensional quantities, it is necessary to implement those modifications in a 3-dimensional air-shower generator such as \corsika~\cite{Heck:1998vt}. While the modification of the cross section is straightforward, as it is applied before the interaction, modifications of other parameters require resampling of secondary particles produced by the hadronic interaction model. The existing resampling code used in~\cite{Ulrich:2010rg} has been extensively tested and is available to the community as part of the standalone \conex package. The \conex code can be used for the simulation of the first part of the shower development in \corsika under the \conexD option~\cite{conex3D}. In this option, the particles are eventually handed over to standard \corsika upon crossing configurable energy thresholds, so that the full 3-dimensional shower is recreated. With small adjustments, we have ported the available resampling code for use under the \conexD option in \corsika so that full 3-dimensional showers can be generated with modified hadronic interactions.

The implementation in \corsika addresses two limitations of the original resampling code: the inability to perform resampling of different variables simultaneously and the universality of $E_{\mathrm{thr}}$ values for all modifications. 
On the other hand, we have opted not to implement modifications of the pion charge ratio. While this was proven in \cite{Ulrich:2010rg} to be an efficient way increasing the muon content of air showers, many other modifications for the same purpose have been developed since -- most are reviewed in \cite{Albrecht:2021cxw} except for  the latest works such as  \cite{RIEHN2024102964} -- and thus implementing a single mechanism would be somewhat arbitrary.
There is now an abundance of data on the production of specific mesons at high energies, from precise measurements of neutral pions at the LHC~\cite{ALICE:2017nce} to fixed-target measurements of Kaons at NA61~\cite{NA61SHINE:2023azp}, which actually indicates a violation of isospin symmetry -- any modification to particle production would have to take now into account the rather complex constraints from these measurements, which is beyond the scope of our work. For all of the other modifications, additional steering cards were implemented using the standard \corsika format, making it possible to specify the values $E_{\mathrm{thr}}$ and $f_{19}$ independently for cross section, elasticity, and multiplicity when setting up a \corsika simulation.

\begin{figure}
    \centering
    \includegraphics[width=\linewidth]{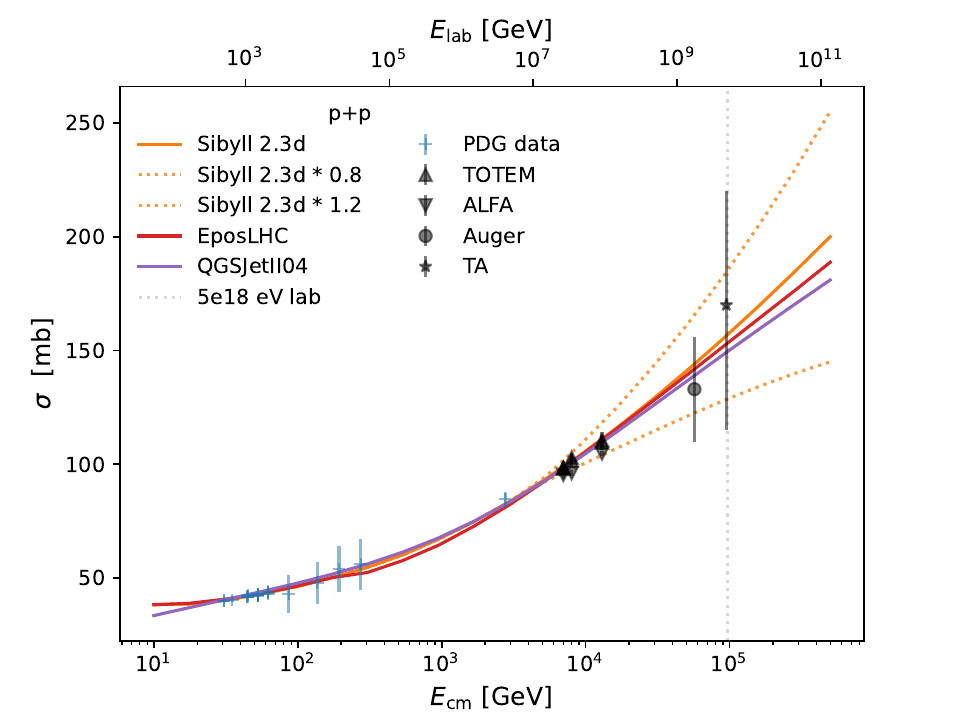}
    \caption{Total cross section in a p-p interaction as a function of CMS and laboratory collision energy predicted by various hadronic interaction models. The \sib{2.3d} model is shown together with modifications utilized in  \cite{icrc2021}. The data points are taken from several experiments, combining statistical and systematic uncertainty in quadrature. The lower energy data (blue crosses) were compiled from~\cite{PDG}. Black upper triangles represent the results of TOTEM~\cite{totem_cross_section}. Black lower triangles represent results from the ALFA detector~\cite{ALFA2016},~\cite{ALFA2022}. The most energetic points show the results obtained by Pierre Auger Collaboration~\cite{PierreAuger:CS} and the Telescope Array~\cite{TA:CS}. The grey line represents the energy used for validation between the different frameworks in this work.}
    \label{fig:cross_section_f19}
\end{figure}

\begin{figure}
    \centering
    \includegraphics[width=\linewidth]{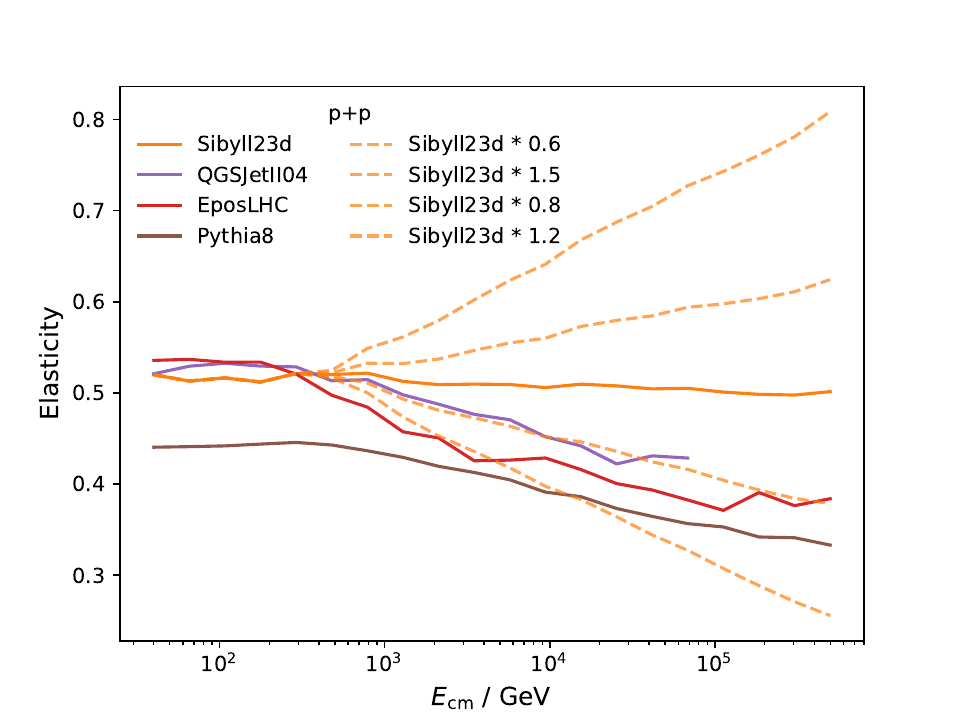}
    \caption{Evolution of elasticity for various hadronic interaction models for a proton-proton interaction. The \sib{2.3d} model is shown together with modifications utilized in \cite{icrc2021}.}
    \label{fig:elasticity_f19}
\end{figure}

\begin{figure}
    \centering
    \includegraphics[width=\linewidth]{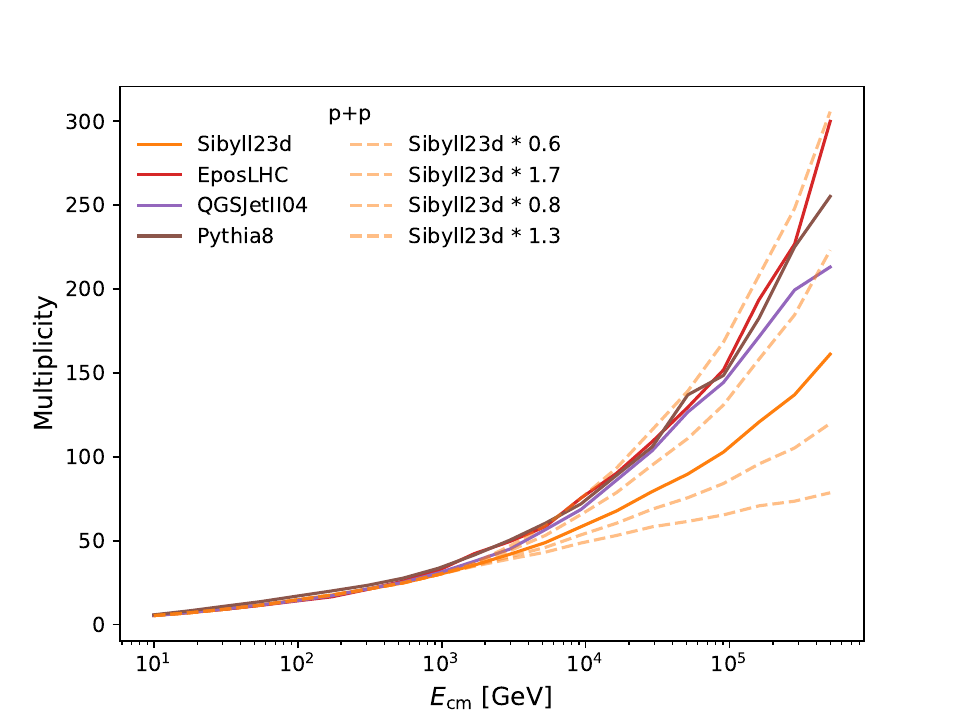}
    \caption{Evolution of multiplicity of charged particles for various hadronic interaction models for a proton-proton interaction. The \sib{2.3d} model is shown together with modifications utilized in \cite{icrc2021}.}
    \label{fig:multiplicity_f19}
\end{figure}

The \conex code for modified simulations is capable of treating nuclear projectiles while using Sibyll \cite{Engel:2019dsg} as the high-energy interaction model. As Sibyll is based on the superposition model of nuclear interactions, each nucleus-air interaction is treated simply as a set of nucleon-air interactions at energy $E/A$. For technical reasons, we have changed the implementation of the modification of the cross section of nuclear projectiles, where we replace the rather complex original calculation (which is difficult to implement under \corsika) with a parameterization that does not produce significant differences for the longitudinal shower development (see \cite{uhecr2022} for details). Due to the energy $E$ in Eq. (2) being taken per nucleon in the case of nuclear projectiles, the effects of modifications are much smaller for nuclear primaries (and equal to those for proton-air collisions at $E/A$). Thus we present results mostly for proton primaries, where the effects are more sizable.

\section{Validation of the modified simulations}

In their original work, the authors of \cite{Ulrich:2010rg} have set $E_{\mathrm{thr}}=10^{15}$ eV for all modifications, above the maximum LHC energy at that time. On the one hand, since the modification factor only grows logarithmically with energy and there are various uncertainties in the experimental constraints for the modified variables, in particular for proton-air or even pion-air interactions, it is reasonable to investigate the effect of at least some of the modifications for thresholds even below the maximal accelerator energy (at least for modest values of $f_{19}$) -- it can in fact be beneficial to set it as low as possible so that we avoid postulating very sharp changes in the properties of the interactions. On the other hand, the maximum accelerator energy has increased by almost two orders of magnitude in the laboratory frame over time. The possible choices of thresholds are thus different for the different variables.

This point is better illustrated by comparing the predictions of the different hadronic interaction models (\sib{2.3d}, \qgsii \cite{qgsjet}, \eposlhc \cite{epos-lhc} and Pythia 8 \cite{Pythia}) for the different variables, as it is shown in Figures \ref{fig:cross_section_f19}, \ref{fig:elasticity_f19} and \ref{fig:multiplicity_f19}, generated using the CHROMO~\cite{chromo} package. While cross-section measurements are directly available, the constraints on elasticity and multiplicity are derived indirectly from other measurements and thus experimental data are shown only in Fig.~\ref{fig:cross_section_f19}. These data \cite{PDG,atlas_cross_section, totem_cross_section, ALFA2022} present strong constraints, making it difficult to justify any choice of $E_{\mathrm{thr}}<10^{16}$~eV; the Auger \cite{PierreAuger:CS} and Telescope Array \cite{TA:CS} measurements at higher energies come with large uncertainties. For elasticity, due to the lack of precise forward measurements, already the current models show large differences in predictions across a wide range of energies; for technical reasons however, we do not consider values lower than  $E_{\mathrm{thr}}=10^{14}$~eV.  For multiplicity, the current models seem to diverge around $E_{\mathrm{thr}}=10^{15}$~eV. In all three figures, we also show how modifications for the respective choices of $E_{\mathrm{thr}}$ for each variable grow above the threshold according to Eq.~\ref{eq_f19} for various choices of $f_{19}$ considered in  \cite{icrc2021}.

In \conex, particles are first treated using the Monte Carlo method, until they reach an energy threshold $E_{\mathrm{CE}}$ below which they are no longer followed individually, but are used as initial conditions for a set of cascade equations. The resampling code only affects the Monte Carlo part of the simulation and thus the threshold energy for cascade equations $E_{\mathrm{CE}}$ must always be set lower than the lowest $E_{\mathrm{thr}}$ for the used modifications. The \conexD option and its consistency with standard \corsika simulations for primary energies $E_0$ up to $10^{19}$ eV have been tested in \cite{Conex3D_2025}. However, the study was carried out under the default configuration, where for hadrons, $E_{\mathrm{CE}}=10^{-3}E_0$, which is significantly higher than the range of \conex Monte Carlo needed to simulate UHECR showers with $E_{\mathrm{thr}}$ for modifications of the order of $10^{14}$ eV. Furthermore, the published tests do not include a comparison of lateral and longitudinal muon profiles, nor do they concern any modified simulations or study the consistency in fluctuations of observables, as they are carried out using small numbers of simulated showers. We thus need to present an extensive validation of the framework for the modified simulations before it can be applied to any physical problems. 

We divide the consistency checks into three categories. First, compare the \textbf{longitudinal} profiles between the three frameworks: \conex, \conexD and \corsika. Second, we look in detail at the shapes of the \textbf{lateral} profiles obtained from \corsika and \conexD (standalone \conex does not provide lateral profiles). Third, we compare the effects of \textbf{modifications} between the \conex and \conexD frameworks (the modifications are not implemented in \corsika without \conex).

\subsection{Simulation setup}

In all simulations, we use \corsika 7.7410 and \conex 7.60 with Sibyll 2.3d \cite{Engel:2019dsg} and UrQMD 1.3.1 \cite{Bleicher_1999} hadronic interaction models, with the energy threshold for the transition between the high and low energy models set to 80 GeV. The primary energy is set to $E_0=5$ EeV, and the primary particles are protons and iron nuclei. As our main goal is to validate the \conexD simulations under conditions allowing the most extensive feasible modifications, we set the energy for the transition between the \conex Monte Carlo part and the cascade equations at $E_{\mathrm{CE}}=2\times 10^{-6} E_0 =10^{13}$ eV for hadrons in \conexD. To avoid introducing unnecessary differences into the comparison, we use the same value for 1-dimensional \conex; the threshold for electromagnetic particles is kept at the default $10^{-4}E_0$). The thresholds for the handover of particles back to \corsika in \conexD simulations were kept at default values (300 GeV for hadrons, 1 GeV for EM particles; muons are always passed to \corsika upon production). For consistency, we used the same parametrization of atmosphere for all three simulation packages, namely the ``Malargüe GDAS model for May after Will/Keilhauer" (MODATM = 22 in the \corsika steering input), which we parameterized for use within \conex.

\begin{figure}
    \centering
    \includegraphics[width=\linewidth]{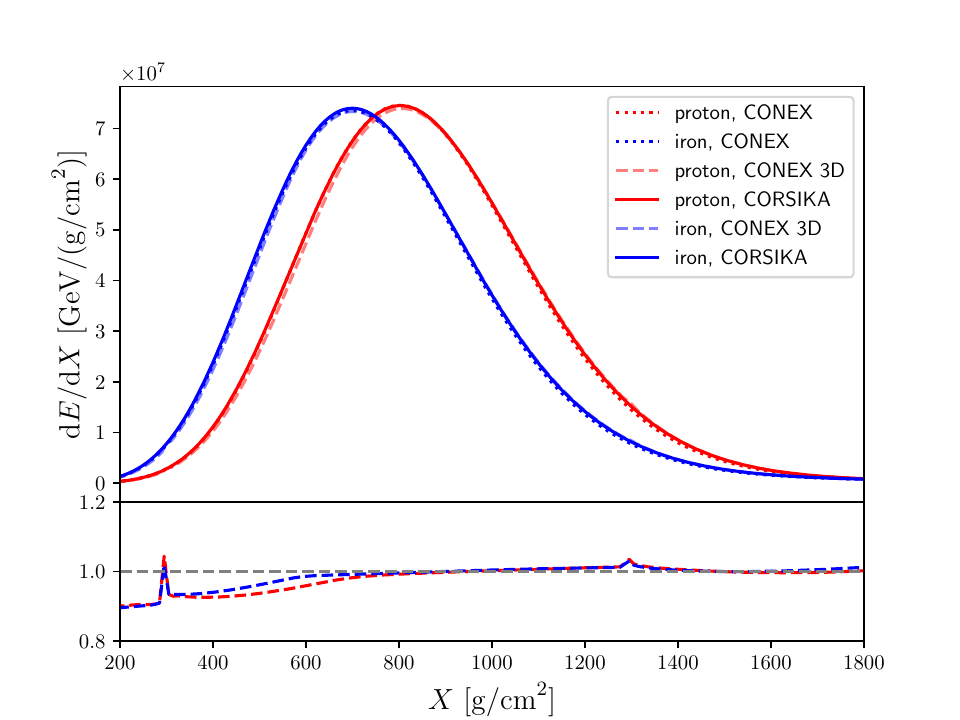}
    \caption{Average longitudinal profiles of energy deposit for iron and proton primaries simulated with \conex, \corsika and \conexD and the ratio between the average profiles for \conexD and \corsika. In the upper part of the plot, the dotted line (\conex) overlaps with the dashed line (\conexD).}
    \label{fig:validation_dEdX_mean}
\end{figure}

\begin{figure}
    \includegraphics[width=\linewidth]{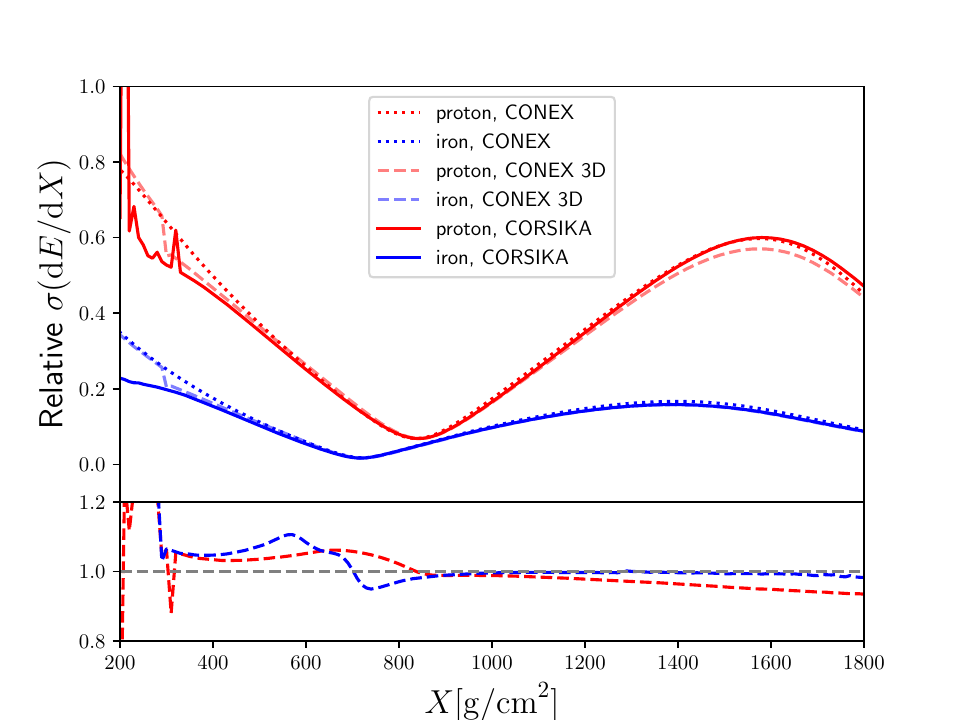}
    \caption{Relative fluctuations of energy deposit for iron and proton primaries simulated with \conex, \corsika and \conexD and the ratios between them for \conexD and \corsika.}
    \label{fig:validation_dEdX_variance}
\end{figure}

\begin{figure}
    \includegraphics[width=\linewidth]{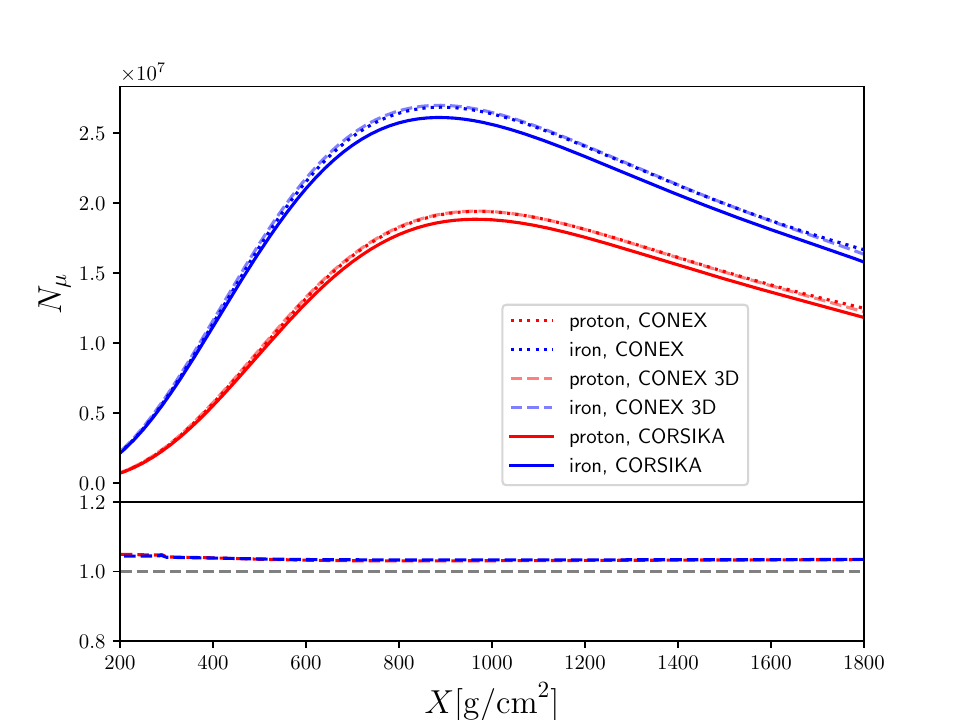}
    \caption{Average longitudinal profiles of the number of muons above 1 GeV for iron and proton primaries simulated with \conex, \corsika and \conexD and the ratios between them for \conexD and \corsika.}
    \label{fig:validation_muons_mean}
\end{figure}

\begin{figure}
    \includegraphics[width=\linewidth]{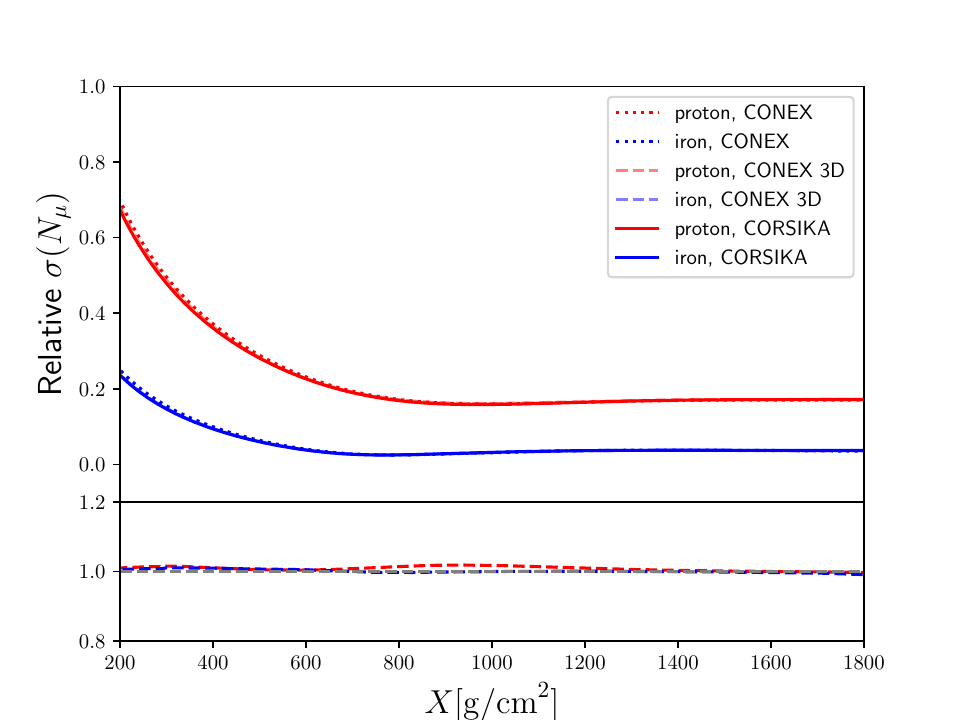}
    \caption{Relative fluctuations of  the number of muons above 1 GeV for iron and proton primaries simulated with \conex, \corsika and \conexD and the ratios between them for \conexD and \corsika.}
    \label{fig:validation_muons_variance}
\end{figure}

\subsection{Longitudinal shower development} \label{sec:long}

\begin{table}
\caption{Comparison of mean and standard deviation values of $X_{\max}$ across simulation frameworks for showers at primary energy 5 EeV simulated using Sibyll~2.3d hadronic interaction model.}
  \centering
  \begin{ruledtabular}
   \begin{tabular}{ccccc}
     & \multicolumn{2}{c}{proton} &
        \multicolumn{2}{c}{iron} \\
     & $\boldsymbol{\langle X_{\max}\rangle}$  & 
       $\boldsymbol{\sigma (X_{\max})}$ & 
       $\boldsymbol{\langle X_{\max}\rangle}$  & 
       $\boldsymbol{\sigma (X_{\max})}$
      \\
     & [g/cm$^{2}$]& [g/cm$^{2}$]& [g/cm$^{2}$]& [g/cm$^{2}$]\\
         \colrule
     \conex & $797.8\pm0.2$ & $62.0\pm0.2$ & $695.0\pm0.1$ & $22.8\pm0.1$ \\ 
     \conexD & $798.1\pm0.8$ & $63.7\pm0.5$ & $694.2\pm0.3$ & $23.3\pm0.2$ \\ 
     \corsika & $797.7\pm0.7$ & $62.4\pm0.5$ & $694.9\pm0.3$ & $23.0\pm0.2$ \\
  \end{tabular}
  \end{ruledtabular}
  \label{tab:xmax_comparison}
\end{table}

\begin{figure*}
    \centering

    \begin{subfigure}{\textwidth}
        \centering
        \includegraphics[width=.48\textwidth,height=\latfigheight,keepaspectratio]{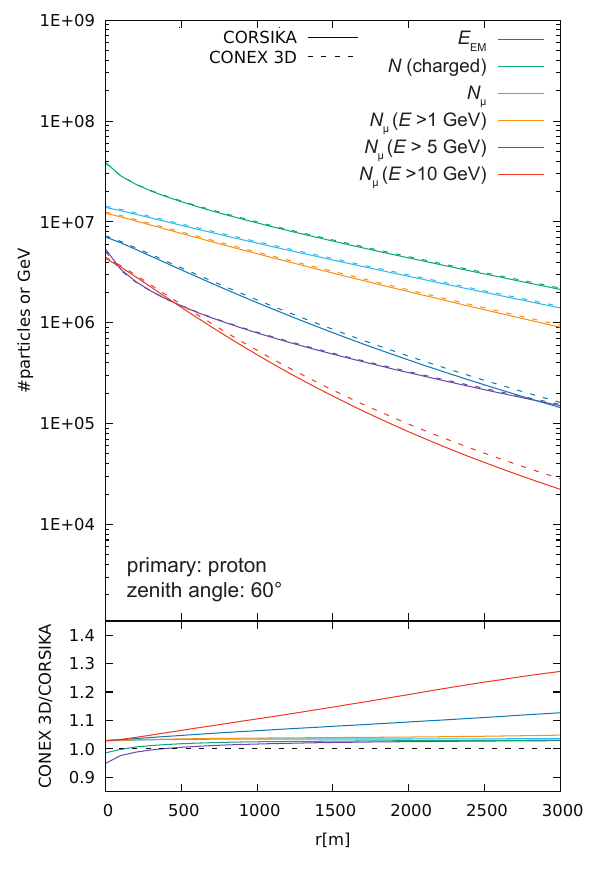}
        \hspace{0.01\textwidth}
        \includegraphics[width=.48\textwidth,height=\latfigheight,keepaspectratio]{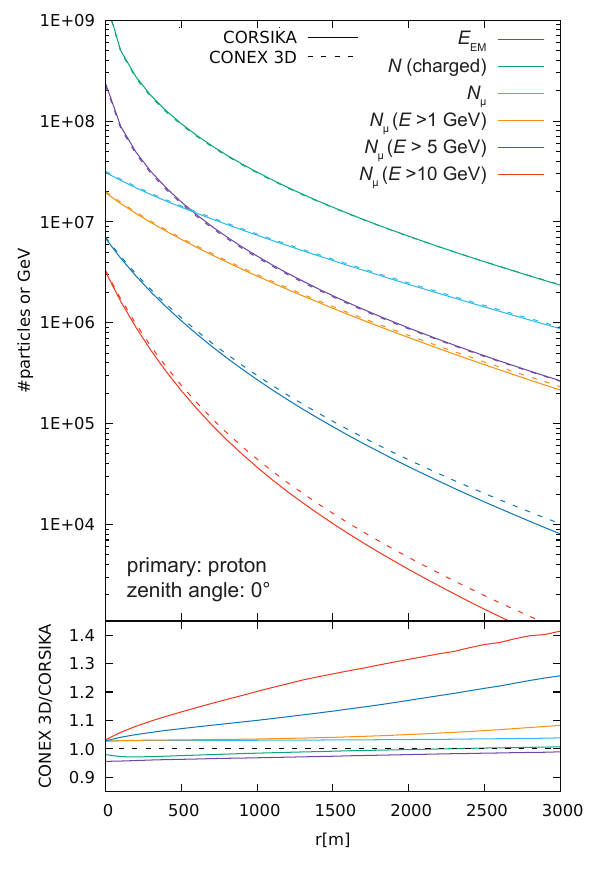}
        \caption{Primary protons.}
        \label{fig:validation_lateral_mean_proton_60_0}
    \end{subfigure}

    \vspace{0.8em}

    \begin{subfigure}{\textwidth}
        \centering
        \includegraphics[width=.48\textwidth,,height=\latfigheight,keepaspectratio]{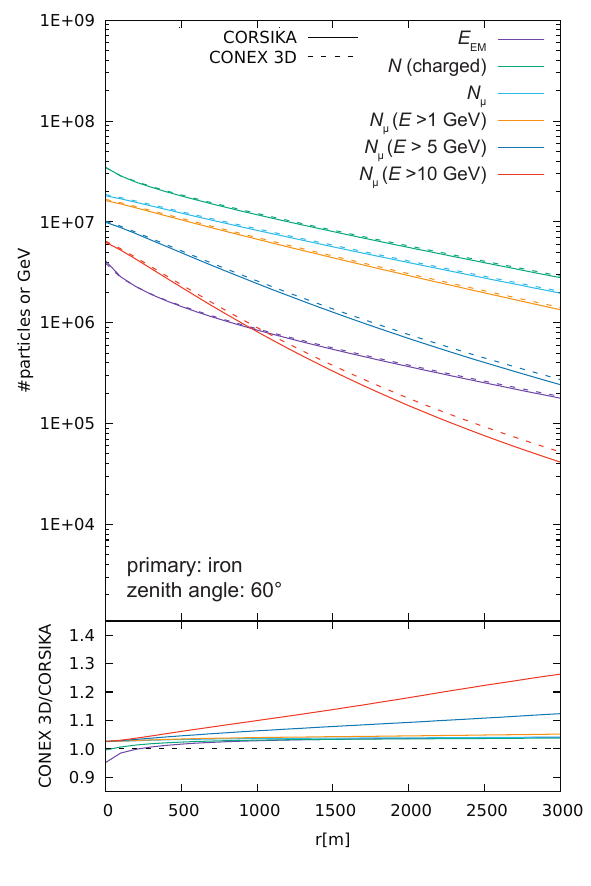}
        \hspace{0.01\textwidth}
        \includegraphics[width=.48\textwidth,height=\latfigheight,keepaspectratio]{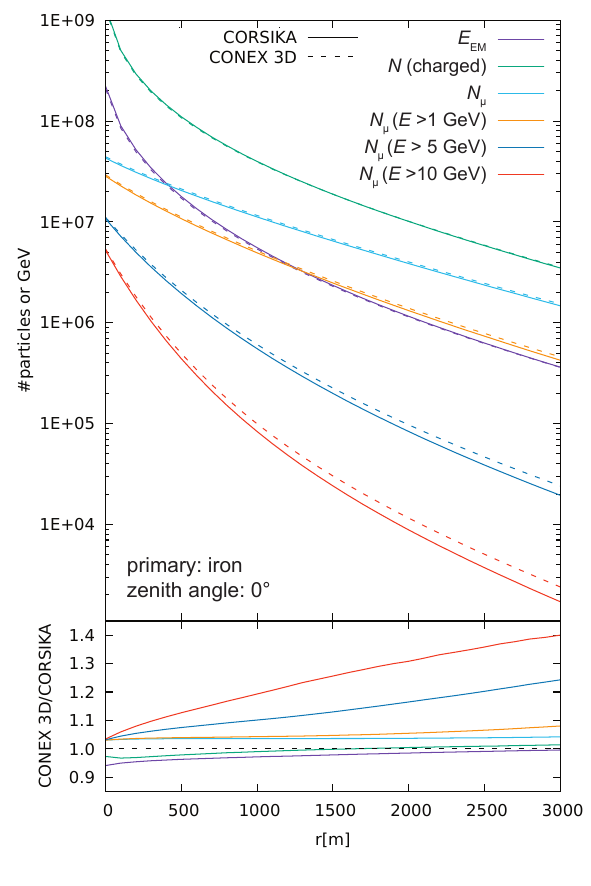}
        \caption{Primary iron nuclei.}
        \label{fig:validation_lateral_mean_iron_60_0}
    \end{subfigure}

    \caption{Mean cumulative (summed over radii from the shower axis $>r$) lateral distributions of different variables for \conexD and \corsika simulations and their ratios, for protons (panel (a)) and iron nuclei (panel (b)). In each panel, the left and right plots correspond to zenith angles $60^\circ$ and $0^\circ$, respectively.}
    \label{fig:validation_lateral_mean_combined}
\end{figure*}

\begin{figure*}
    \centering

    \begin{subfigure}{\textwidth}
        \centering
        \includegraphics[width=.48\textwidth,height=\latfigheight,keepaspectratio]{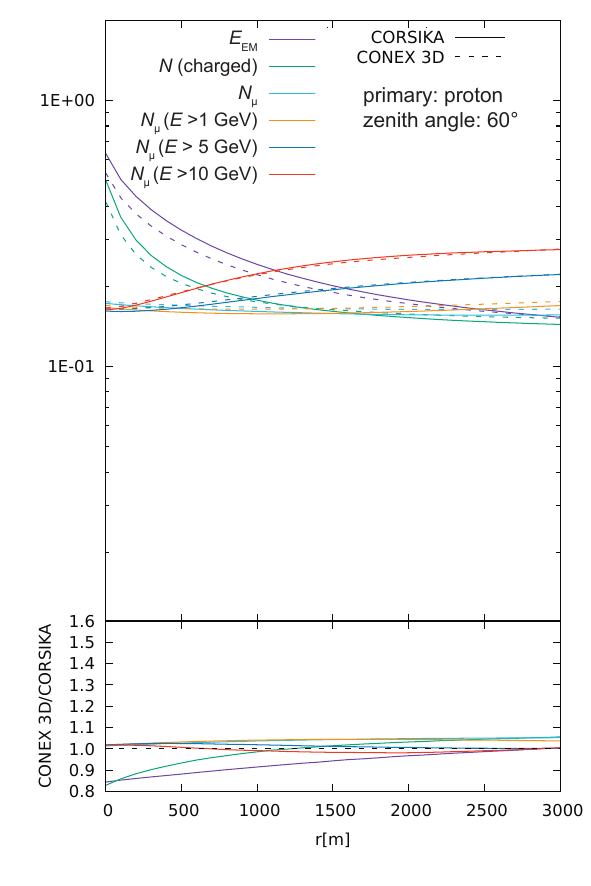}
        \hspace{0.01\textwidth}
        \includegraphics[width=.48\textwidth,height=\latfigheight,keepaspectratio]{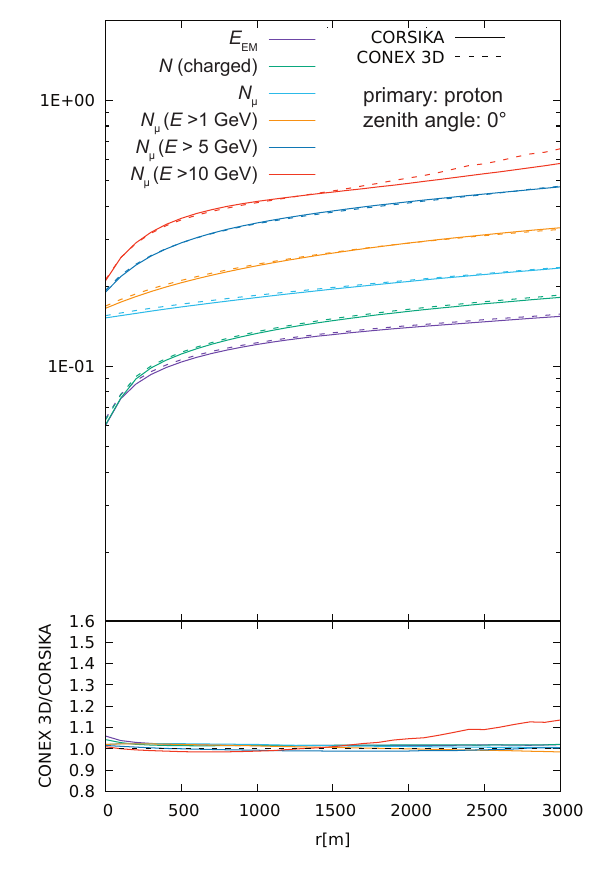}
        \caption{Primary protons.}
        \label{fig:validation_lateral_var_proton_60_0}
    \end{subfigure}

    \vspace{0.8em}

    \begin{subfigure}{\textwidth}
        \centering
        \includegraphics[width=.48\textwidth,height=\latfigheight,keepaspectratio]{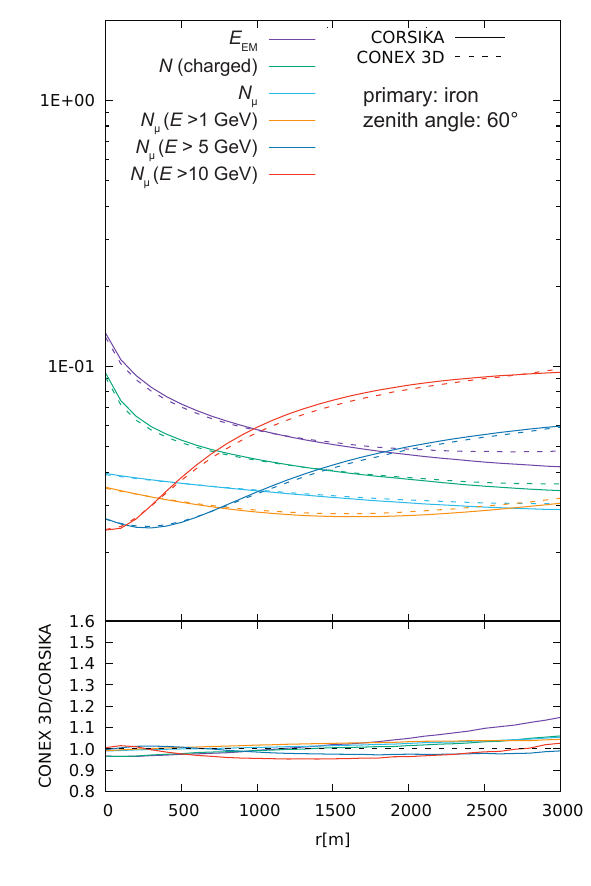}
        \hspace{0.01\textwidth}
        \includegraphics[width=.48\textwidth,height=\latfigheight,keepaspectratio]{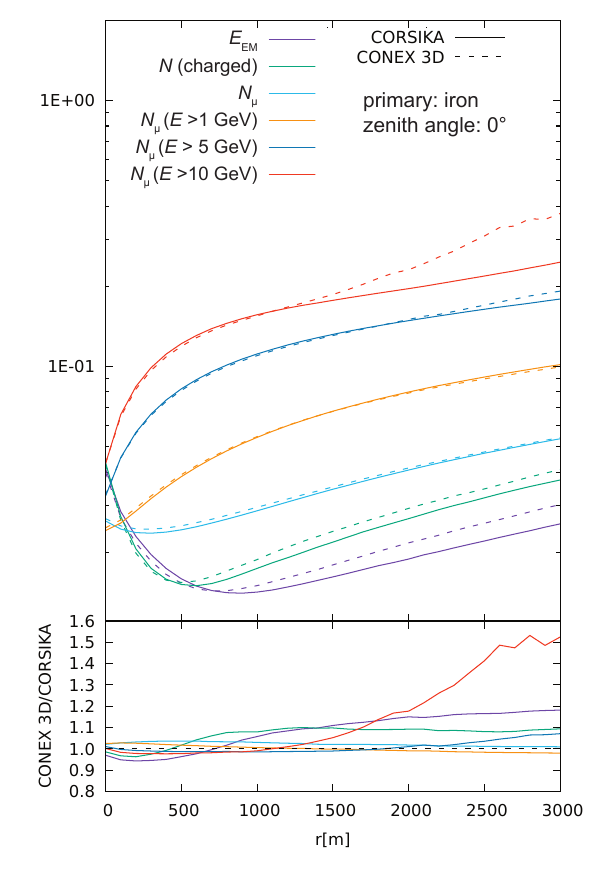}
        \caption{Primary iron nuclei.}
        \label{fig:validation_lateral_var_iron_60_0}
    \end{subfigure}

    \caption{Relative fluctuations of cumulative (summed over radii from the shower axis $>r$) lateral distributions of different variables for \conexD and \corsika simulations and their ratios, for protons (panel (a)) and iron nuclei (panel (b)). In each panel, the left and right plots correspond to zenith angles $60^\circ$ and $0^\circ$, respectively.}
    \label{fig:validation_lateral_var_combined}
\end{figure*}

Longitudinal profiles allow for an efficient comparison between the predictions of the different frameworks across a wide range of slant depths, which effectively corresponds to a large range of combinations of zenith angle and altitude above sea level. However, in the case of inclined showers (which maximize the span of slant depths), care must be taken near the ground in \corsika (and thus also in \conexD) as the longitudinal profile is affected when particles far from the shower axis encounter the ground before the shower axis does. To this end, we use simulations in \corsika and \conexD with the ground altitude set to 0 meters and the zenith angle set to 60 degrees, which provide longitudinal profiles unaffected by this effect for up to 1800 g/cm$^2$ of depth. The energy cuts for particle tracking in these simulations are set to the \conex defaults -- 1 GeV for hadrons and muons and 1 MeV for electrons and photons -- so that comparison with \conex (where these cuts cannot be easily changed) is possible in absolute numbers.

The mean depths of the maxima of the longitudinal profiles of the energy deposit ($\langle X_{\text{max}}\rangle$) agree very well across all three modes of simulation (\conex, \conexD and \corsika), for both proton and iron primaries (see Table~\ref{tab:xmax_comparison}). We note the available statistics: 7000 showers per primary particle were simulated for \corsika and \conexD and 100 000 showers for \conex. Similarly, we observe no incompatibility in $\langle X_{\text{max}}\rangle$ for the iron primaries. When the profiles are compared with each other, they show a difference of up to 10 \% in the early stages between the \corsika and \conex-based frameworks, see Fig. \ref{fig:validation_dEdX_mean}. The structure around 300 g/cm$^2$ seems to be caused by the transition from the Monte Carlo treatment to the cascade equations. A rough calculation based on the average expected multiplicities in hadronic showers (see \cite{MONTANUS20144}) and the expected interaction lengths suggests that the slant depth of 300 g/cm$^2$ corresponds approximately to $\log(E/\mathrm{eV}) = 11.5$ of the mean energy of a hadronic particle present in the cascade, which in turn corresponds to the point where the source terms of the cascade equations are calculated. As it is clear already from Table~\ref{tab:xmax_comparison}, this feature is far from the shower maximum and does not influence the determination of \Xmax from the simulations. The small kink around 1300 g/cm$^2$ may be related to the forced handover of particles to \corsika in \conexD at a specific minimal distance to the ground. 

The standard deviations of \Xmax agree very well across the respective frameworks. The longitudinal profiles of relative fluctuations (deviations divided by the mean) in the energy deposit (Fig.~\ref{fig:validation_dEdX_variance}) show a richer structure, primarily due to fluctuations in \corsika simulations. An interesting trend is visible where the relative fluctuations in \conexD profiles start following that of \conex, but around 300 g/cm$^2$ abruptly turn closer to the \corsika line. The longitudinal profiles of the number of muons (Fig.~\ref{fig:validation_muons_mean}) and their relative fluctuations (Fig.~\ref{fig:validation_muons_variance}) show a consistent but small $\approx$~5\% overestimation by \conexD compared to the \corsika simulations. Since in \conexD the muons are handed immediately over to \corsika upon creation, the difference must come from the differences in the hadronic cascades.

\subsection{Particles at ground}  \label{sec:ground}

To validate the 3-dimensional structure of the \conexD simulations, we compare the lateral distribution of particles at the ground with those obtained from \corsika simulations. Since the \conexD option in \corsika is not compatible with the option for multiple observation levels, we are forced to choose a single ground level, which we set to 1400 meters a.s.l., the altitude of the Pierre Auger Observatory~\cite{PierreAuger:2015eyc}. We use simulations at different zenith angles to investigate the effects of different slant depths. The simulated showers are filtered to ensure the completeness of the longitudinal profiles and to avoid the occasional incomplete showers encountered in mass production on the computing cluster. Since the lateral comparison cannot involve \conex, we are free to choose finer particle energy cuts, as is more typical in simulations for UHECR experiments: 300 MeV for hadrons, 10 MeV for muons, and 250 keV for electromagnetic particles.

We extract the energy density in the EM component, the number density of all charged particles and the number densities for muons above different thresholds in 50-meter rings in the projected plane perpendicular to the shower axis. Even with 6--8 thousand simulations per primary particle and zenith angle, the mean lateral distribution functions for muons show large fluctuations between radial bins; as the distributions are steeply falling with radius, we effectively smooth those out by considering cumulative sums above each given radius $r$. Figures \ref{fig:validation_lateral_mean_combined}\subref{fig:validation_lateral_mean_proton_60_0} and \ref{fig:validation_lateral_mean_combined}\subref{fig:validation_lateral_mean_iron_60_0} show that while lateral distributions of EM energy ($E_\text{em}$) and low-energy muons are very well reproduced in \conexD across a large range of radii, there are significant discrepancies for muon energies higher than 5 GeV, in particular, far from the shower axis and for more vertical showers. For relative fluctuations of the same variables, the agreement is again generally good. For proton primaries (Fig.~\ref{fig:validation_lateral_var_proton_60_0}), the main discrepancy is in relative fluctuations for all charged particles and EM energy density at small $r$ and high zenith angle, while for iron nuclei (Fig.~\ref{fig:validation_lateral_var_iron_60_0}), there are discrepancies for higher $r$ and small zenith angle, with additional very large relative fluctuations for high-energy muons in this case. Clearly, the \conexD simulations are more faithful in some settings and less in others, and their applicability should be considered with the specific use case in mind. Table~\ref{tab:ground} provides numerical values of ratios between \conexD and \corsika simulations for various quantities shown in Figs.~\ref{fig:validation_lateral_mean_combined}--\ref{fig:validation_lateral_var_combined}.

\subsection{Reproducibility of modifications}  \label{sec:rep}

Using the same setup as in the case of the comparison of longitudinal profiles in Section~\ref{sec:long}, we have created simulations with the same individual modifications of cross section, multiplicity and elasticity in both \conex and \conexD  to verify that the effects of the simulations on longitudinal quantities remain consistent. For each of the three interaction characteristics, we have chosen the two extreme values of $f_{19}$ used in \cite{icrc2021} and the subsequent publications, along with the respective choices of the energy thresholds,  as these roughly correspond to the maximum allowable modifications given the current experimental constraints. The number of simulated showers for \conex was 10$^5$ events per modification bin, and for \conexD, 7000 events per modification bin.

\begin{figure}
    \centering
    \includegraphics[width=\linewidth]{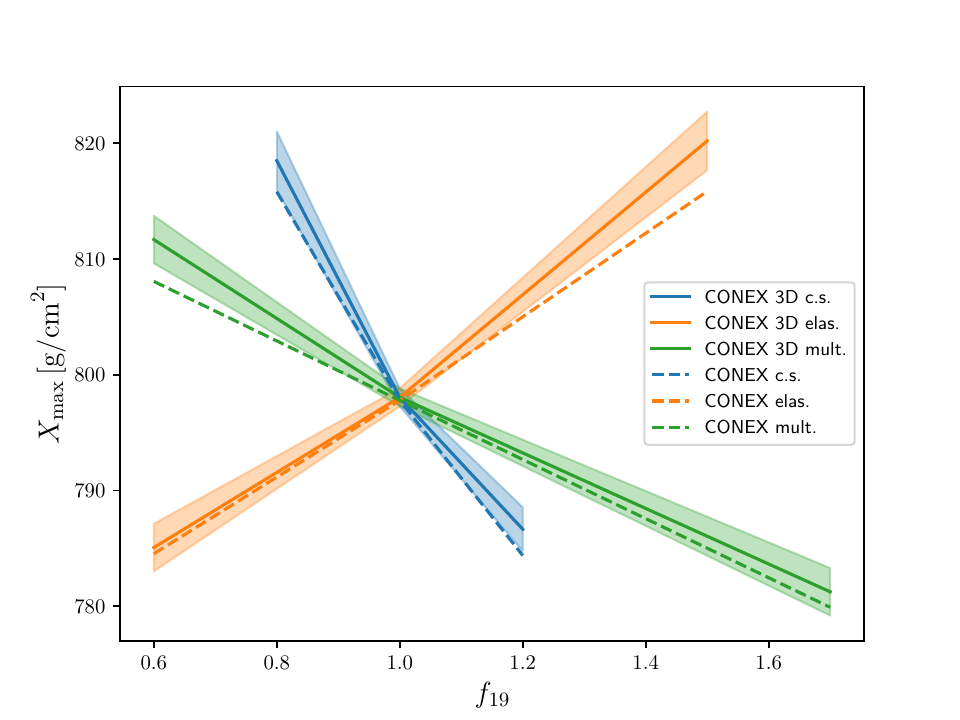}
    \caption{Comparison of \conex and \conexD implementations of the resampling algorithms for the mean \Xmax for primary protons.}
    \label{fig:retarded_octopus_mean_xmax}
\end{figure}

\begin{figure}
    \centering
            \includegraphics[width=\linewidth]{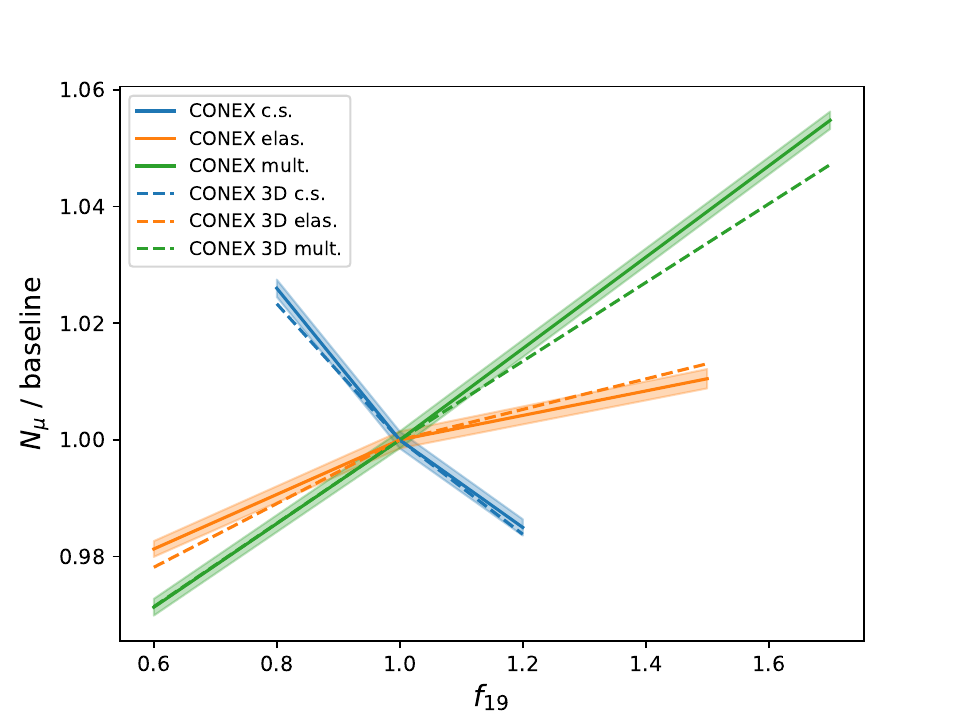}
        \caption{Comparison of \conex and \conexD implementations of the resampling algorithms for the number of muons at $X=1760$~gcm$^{-2}$ for primary protons wrt. baseline ($f_{19}=1$).}
    \label{fig:octopus_muons}
\end{figure}

\begin{figure*}
    \centering
    \begin{subfigure}[b]{0.49\linewidth}
        \centering
        \includegraphics[width=\linewidth]{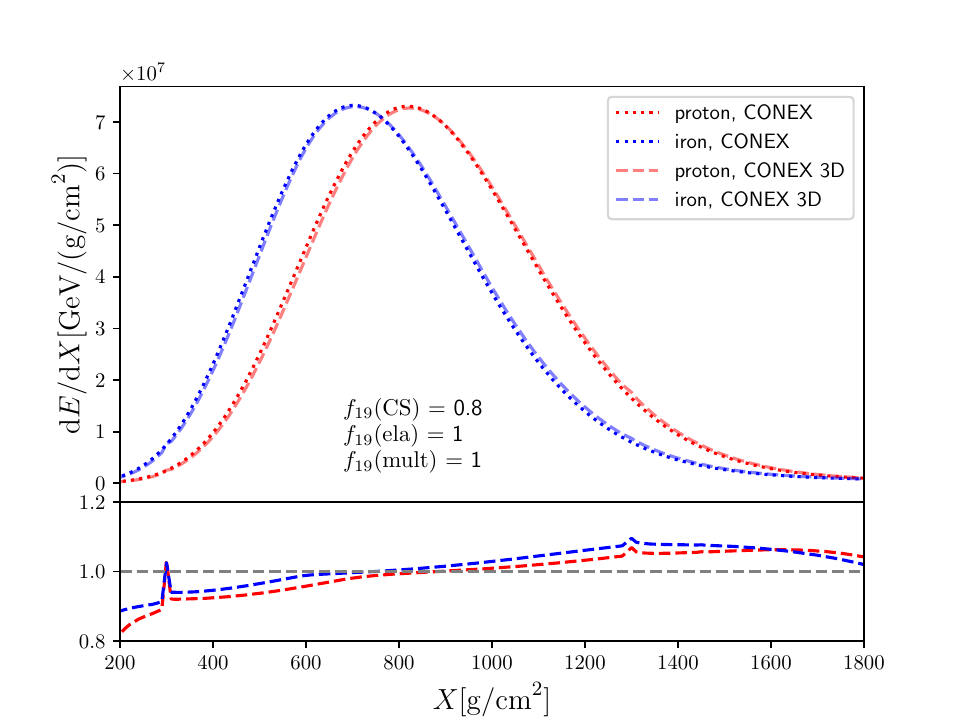}
    \end{subfigure}
    \hfill
    \begin{subfigure}[b]{0.49\linewidth}
        \centering
        \includegraphics[width=\linewidth]{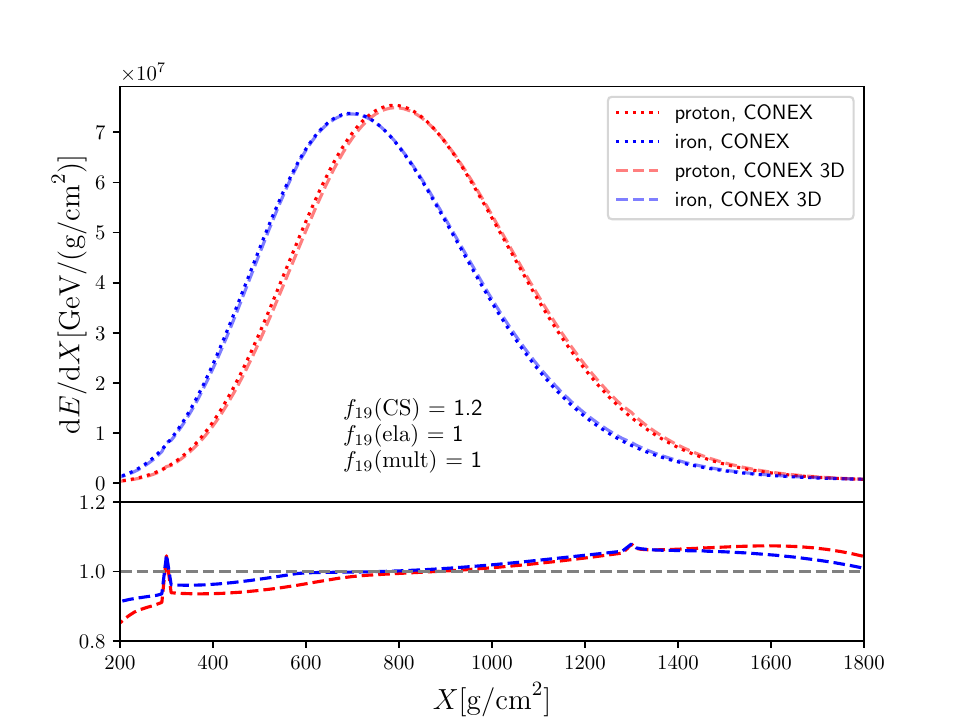}
    \end{subfigure}   
    \vspace{0.5cm}    
    \begin{subfigure}[b]{0.49\linewidth}
        \centering
        \includegraphics[width=\linewidth]{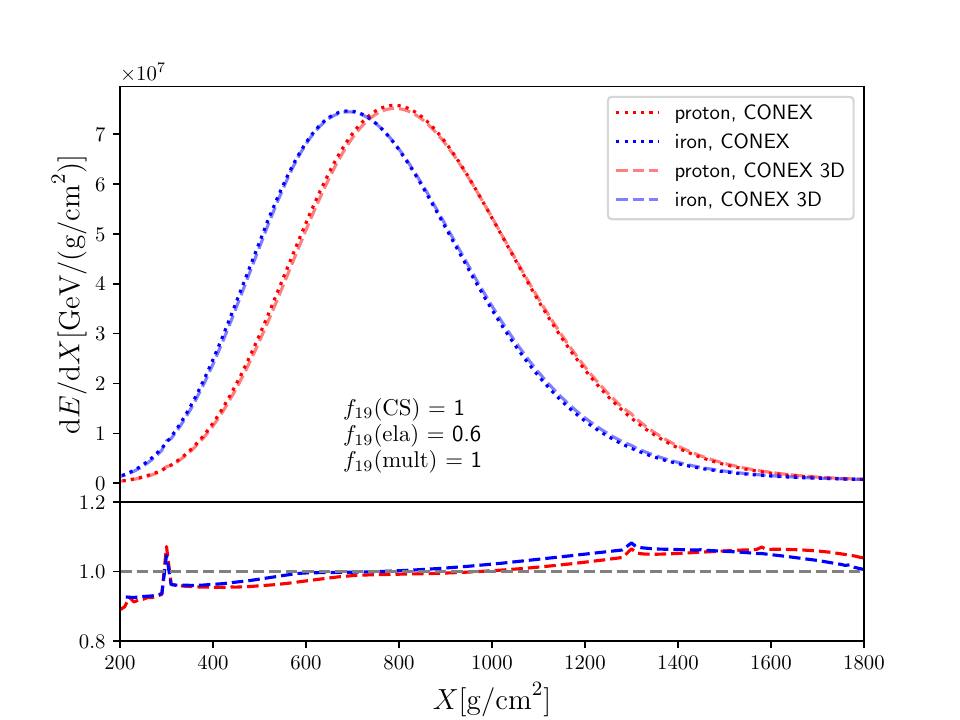} 
    \end{subfigure}
    \hfill
    \begin{subfigure}[b]{0.49\linewidth}
        \centering
        \includegraphics[width=\linewidth]{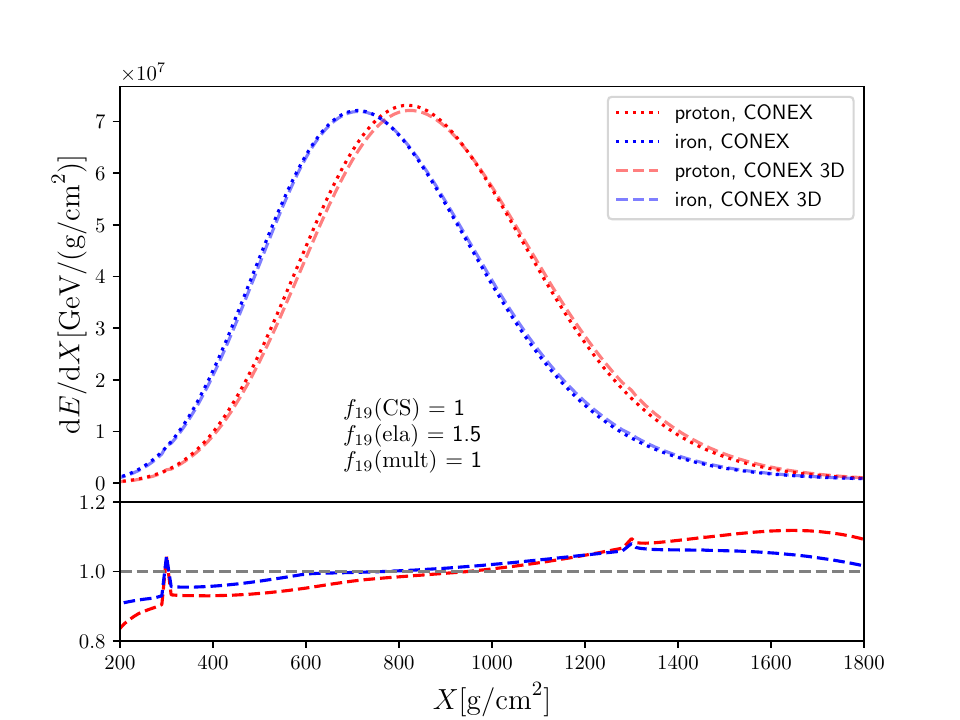}
    \end{subfigure}
    \vspace{0.5cm}
    \begin{subfigure}[b]{0.49\linewidth}
        \centering
        \includegraphics[width=\linewidth]{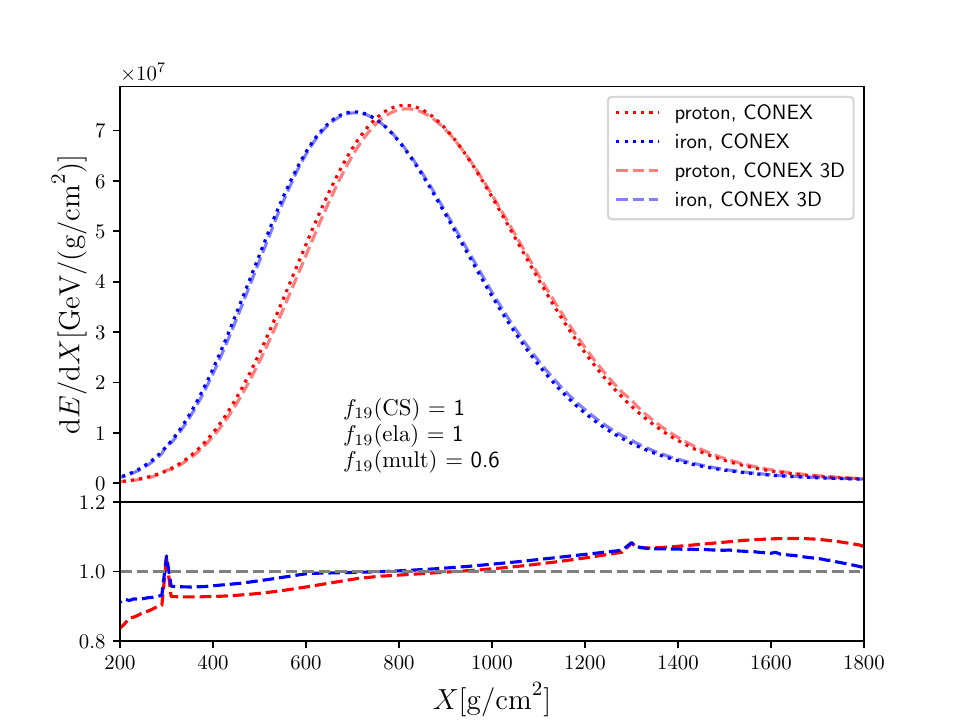} 
    \end{subfigure}
    \hfill
    \begin{subfigure}[b]{0.49\linewidth}
        \centering
        \includegraphics[width=\linewidth]{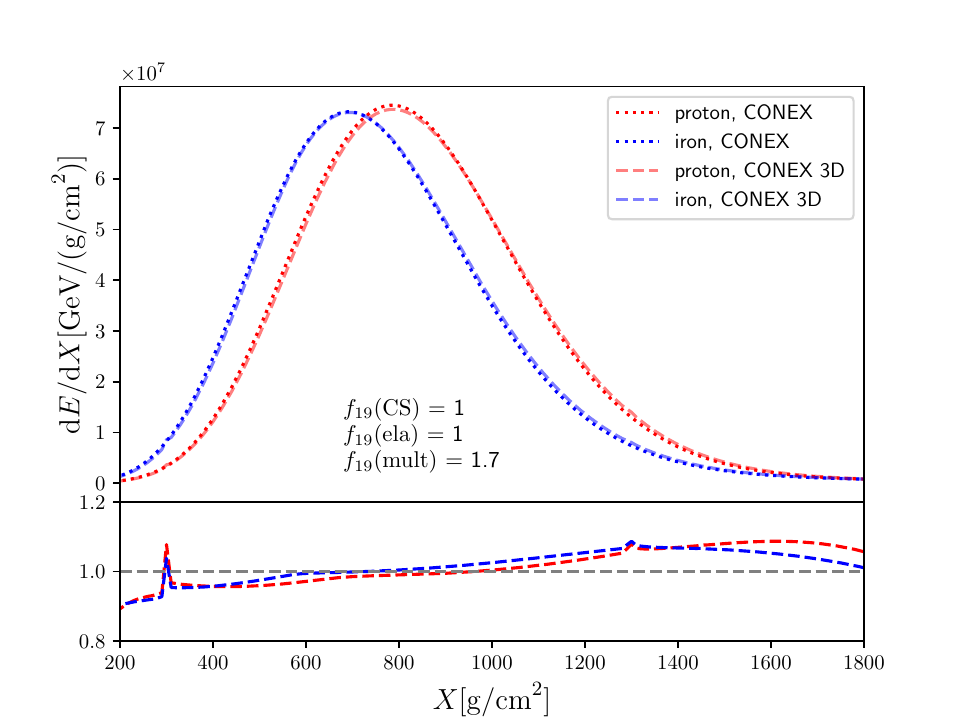}
    \end{subfigure}    
    \caption{Longitudinal profiles of $\mathrm{d}E/\mathrm{d}X$.  The first row corresponds to a modified cross section, the second to a modified elasticity, and the third to a modified multiplicity.}
    \label{fig:validation_parameters_dEdX}
\end{figure*}

\begin{figure*}
    \centering
    \begin{subfigure}[b]{0.49\linewidth}
        \centering
        \includegraphics[width=\linewidth]{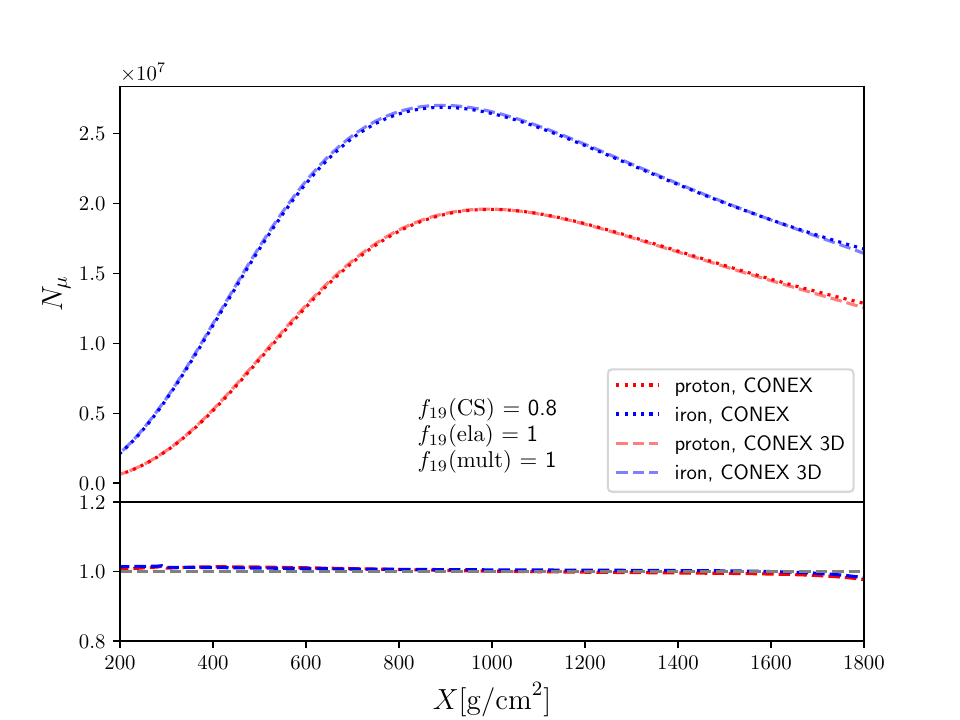}
    \end{subfigure}
    \hfill
    \begin{subfigure}[b]{0.49\linewidth}
        \centering
        \includegraphics[width=\linewidth]{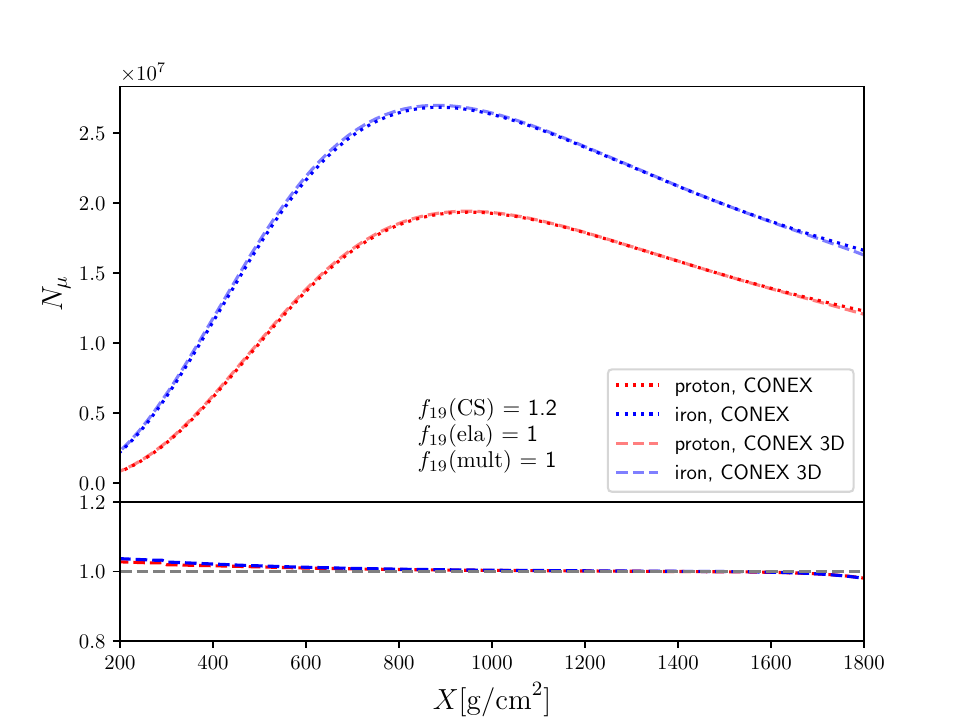}
    \end{subfigure}    
    \vspace{0.5cm}   
    \begin{subfigure}[b]{0.49\linewidth}
        \centering
        \includegraphics[width=\linewidth]{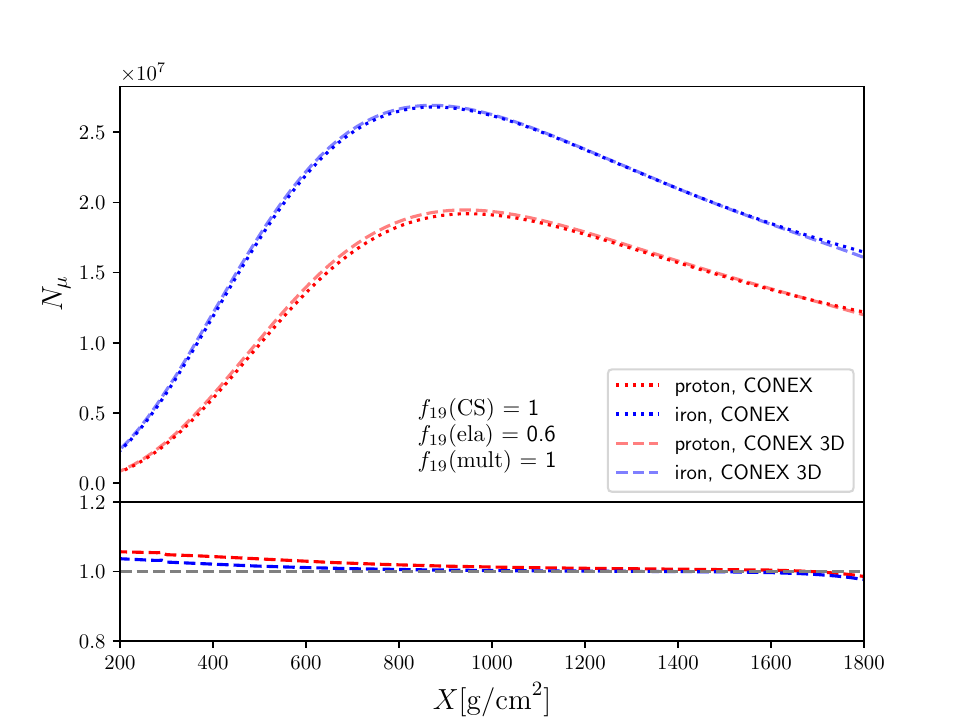} 
    \end{subfigure}
    \hfill
    \begin{subfigure}[b]{0.49\linewidth}
        \centering
        \includegraphics[width=\linewidth]{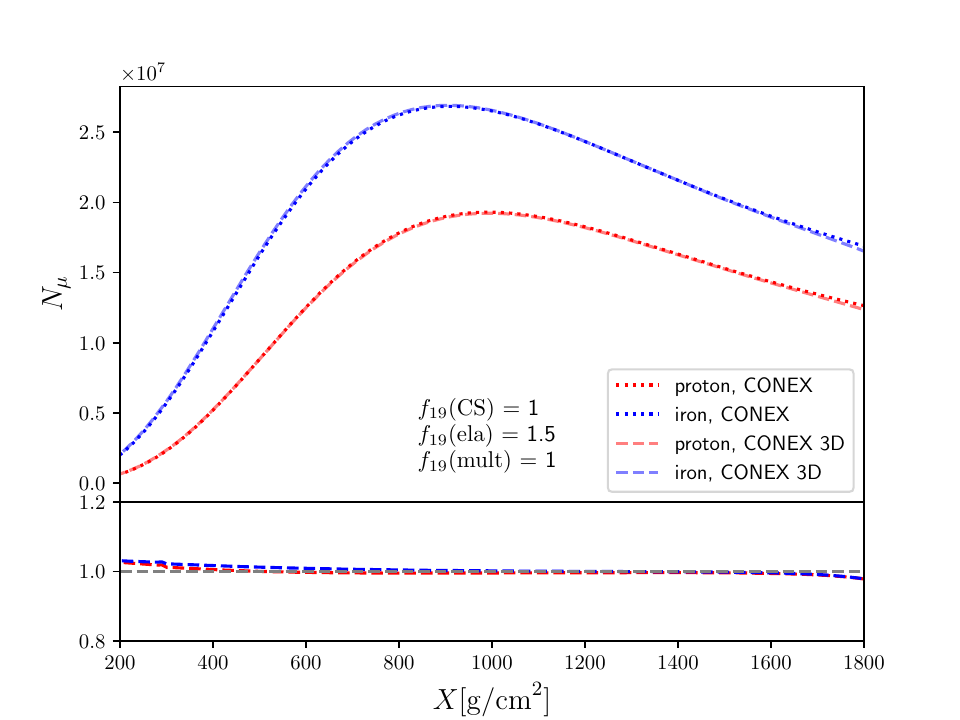}
    \end{subfigure}
    \vspace{0.5cm}
    \begin{subfigure}[b]{0.49\linewidth}
        \centering
        \includegraphics[width=\linewidth]{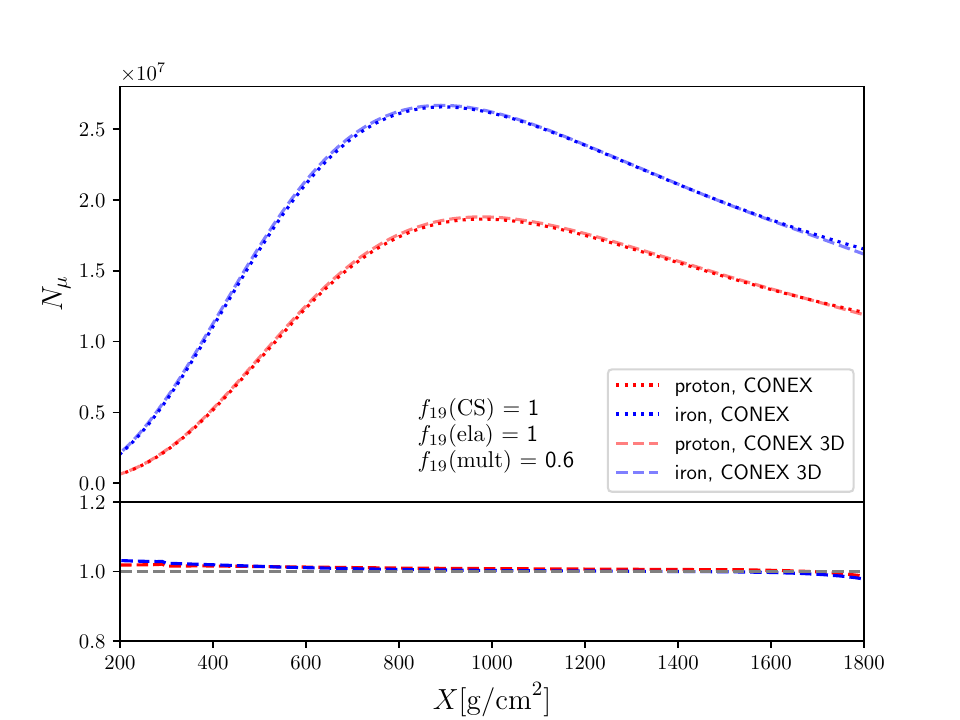} 
    \end{subfigure}
    \hfill
    \begin{subfigure}[b]{0.49\linewidth}
        \centering
        \includegraphics[width=\linewidth]{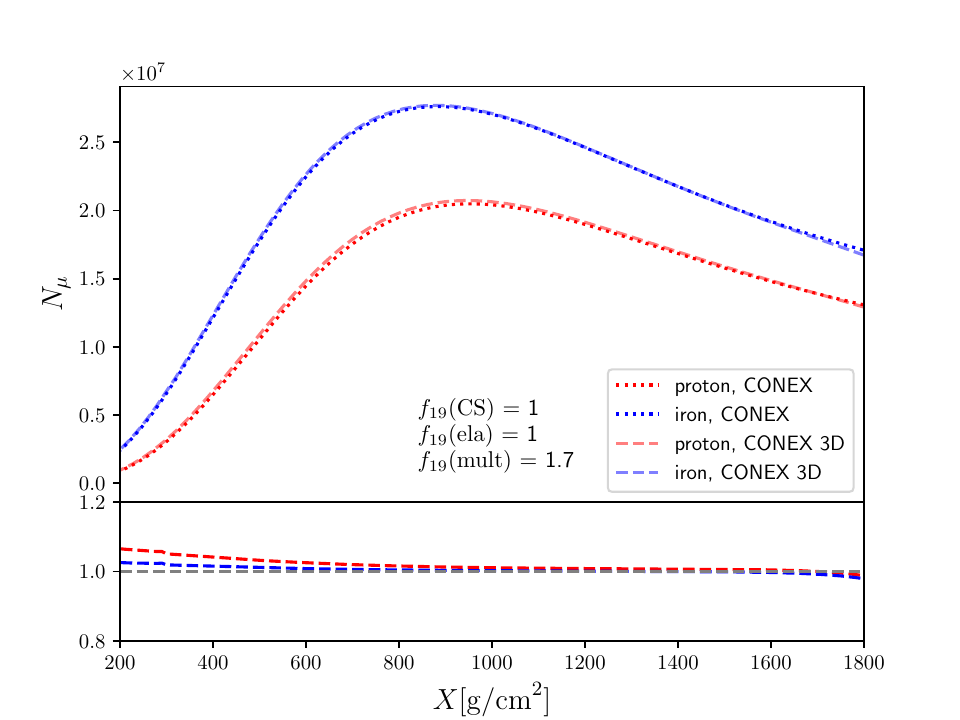}
    \end{subfigure}
        \caption{Longitudinal profiles of the number of muons above 1 GeV. The first row corresponds to a modified cross section, the second to a modified elasticity, and the third to a modified multiplicity.}
    \label{fig:validation_parameters_muons}
\end{figure*}

The comparison can be visualized in two ways. We can plot individual quantities as a function of $f_{19}$ for the three different modifications and both frameworks (Figs.~\ref{fig:retarded_octopus_mean_xmax}  and \ref{fig:octopus_muons}), or  we can compare the longitudinal profiles between the frameworks for the individual modifications (Figs.~\ref{fig:validation_parameters_dEdX} and \ref{fig:validation_parameters_muons}). The first type of comparison shows relatively good agreement within the statistical uncertainties dominated by the much lower statistics of \conexD  simulations -- which, while representing a considerable speed-up compared to equivalent pure \corsika simulations, are still more computationally demanding than \conex simulations -- with the most significant deviation for the number of muons at the chosen level ($X=1760$~gcm$^{-2}$)  and increased multiplicity. A quantitative interpretation of the second type of comparison is more difficult. However, the plots clearly show that the agreement between the longitudinal profiles produced by \conex and \conexD is not significantly affected by the modifications.

\begin{table}
\caption {Ratios between radial cumulative values for various quantities between \conexD and \corsika for proton and iron primary particles at  zenith angles $\theta$ = 0 and 60 degrees. For each quantity, the ratios for mean values and relative fluctuations are shown both for fixed $r=1000$ m and for the maximum deviation from 1.0 across the range of $r=$ to 3000 meters. Deviations from 1.0 larger than 10 \% are highlighted in \textit{italics}, larger than 20 \% in \textbf{bold}.}
\centering
\begin{ruledtabular}
\begin{tabular}{ccccc}
   &
   \multicolumn{2}{c}{mean value} &
   \multicolumn{2}{c}{rel. fluc.} \\
   quantity &
   1000 m &
   max. dev. &
   1000 m &
   max. dev. \\
\colrule  
\multicolumn{5}{c}{proton $\theta = 0$ degrees}\\
\colrule
$E_\mathrm{EM}$ & 0.97 & 0.96 & 1.02 & 1.06 \\
$N_\mathrm{charged}$ & 0.98 & 0.97 & 1.02 & 1.04 \\
$N_\mu$ & 1.03 & 1.04 & 1.02 & 1.02 \\
$N_\mu (E>1$ GeV) & 1.03 & 1.08 & 1.01 & 1.02 \\
$N_\mu (E>5$ GeV) & \textit{1.10} & \textbf{1.26} & 0.99 & 1.02 \\
$N_\mu (E>10$ GeV) & \textbf{1.20} & \textbf{1.41} & 0.99 & \textit{1.14} \\
\colrule
\multicolumn{5}{c}{proton $\theta = 60$ degrees}\\
\colrule
$E_\mathrm{EM}$ & 1.02 & 0.95 & 0.91 & \textit{0.84} \\
$N_\mathrm{charged}$ & 1.02 & 1.03 & 0.99 & \textit{0.83} \\
$N_\mu$ & 1.03 & 1.04 & 1.04 & 1.05 \\
$N_\mu (E>1$ GeV) & 1.04 & 1.05 & 1.04 & 1.04 \\
$N_\mu (E>5$ GeV) & 1.06 & \textit{1.13} & 1.02 & 1.03 \\
$N_\mu (E>10$ GeV) & \textit{1.11} & \textbf{1.27} & 0.99 & 0.98 \\
\colrule
\multicolumn{5}{c}{iron $\theta = 0$ degrees}\\
\colrule
$E_\mathrm{EM}$ & 0.97 & 0.94 & 1.06 & \textit{1.18} \\
$N_\mathrm{charged}$ & 0.99 & 0.97 & 1.08 & 1.10 \\
$N_\mu$ & 1.04 & 1.04 & 1.03 & 1.04 \\
$N_\mu (E>1$ GeV) & 1.04 & 1.08 & 1.01 & 1.03 \\
$N_\mu (E>5$ GeV) & \textit{1.10} & \textbf{1.24} & 0.99 & 1.07 \\
$N_\mu (E>10$ GeV) & \textit{1.19} & \textbf{1.40} & 0.99 & \textbf{1.53} \\
\colrule
\multicolumn{5}{c}{iron $\theta = 60$ degrees}\\
\colrule
$E_\mathrm{EM}$ & 1.03 & 0.95 & 0.99 & \textit{1.15} \\
$N_\mathrm{charged}$ & 1.03 & 1.04 & 0.99 & 1.06 \\
$N_\mu$ & 1.04 & 1.04 & 1.01 & 1.05 \\
$N_\mu (E>1$ GeV) & 1.04 & 1.05 & 1.02 & 1.04 \\
$N_\mu (E>5$ GeV) & 1.06 & \textit{1.12} & 0.98 & 0.97 \\
$N_\mu (E>10$ GeV) & \textit{1.10} & \textbf{1.26} & 0.96 & 0.95 
\end{tabular}
 \end{ruledtabular}
   \label{tab:ground}
\end{table}

\section{Importance of 3D simulations} 

\begin{figure*}
    \centering
    \begin{subfigure}[b]{0.49\linewidth}
        \centering
        \includegraphics[width=\linewidth]{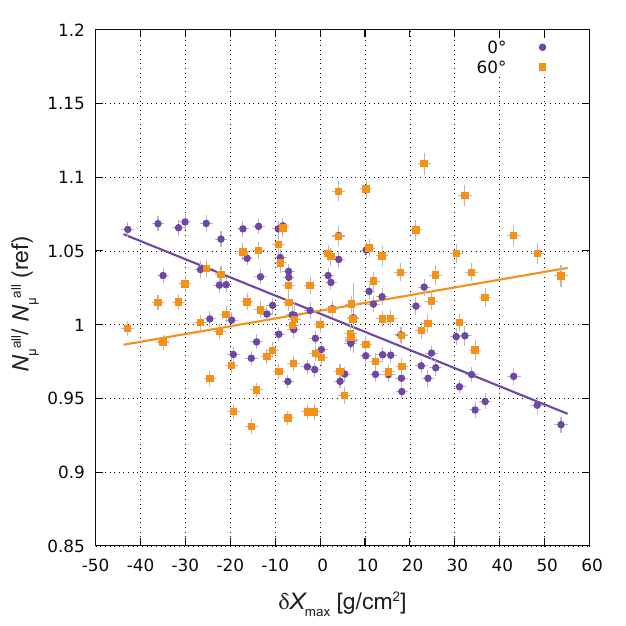}
    \end{subfigure}
    \hfill
    \begin{subfigure}[b]{0.49\linewidth}
        \centering
        \includegraphics[width=\linewidth]{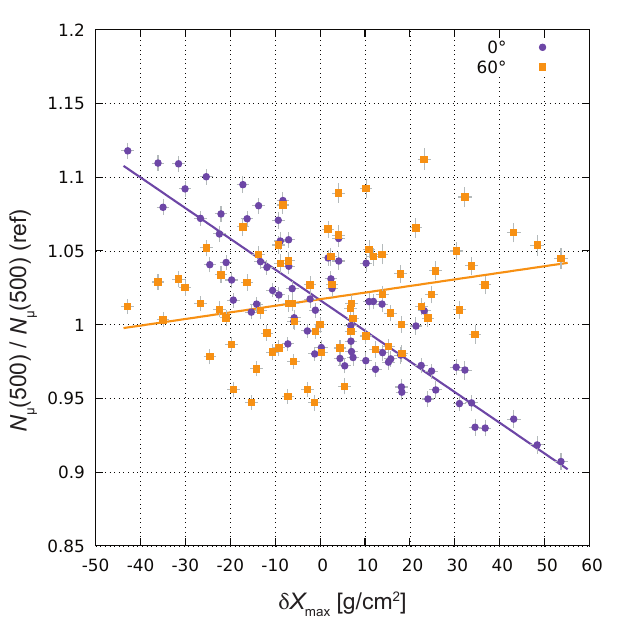}
    \end{subfigure}
    
    \vspace{0.5cm}    
    \begin{subfigure}[b]{0.49\linewidth}
        \centering
        \includegraphics[width=\linewidth]{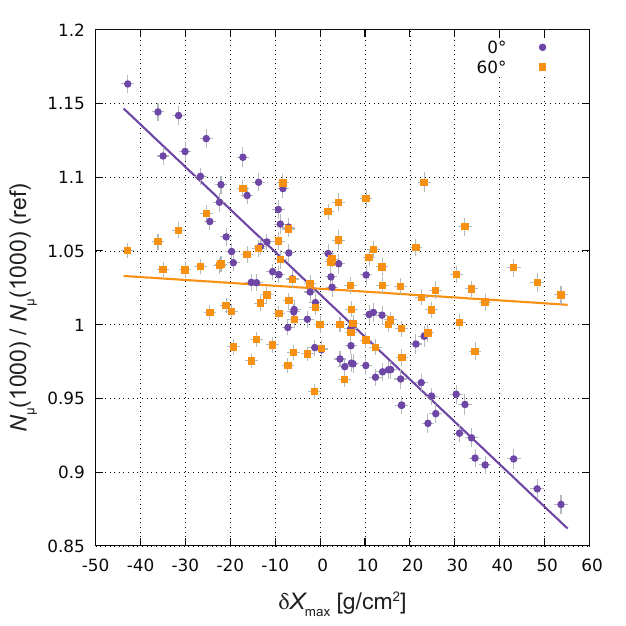} 
    \end{subfigure}
    \hfill
    \begin{subfigure}[b]{0.49\linewidth}
        \centering
        \includegraphics[width=\linewidth]{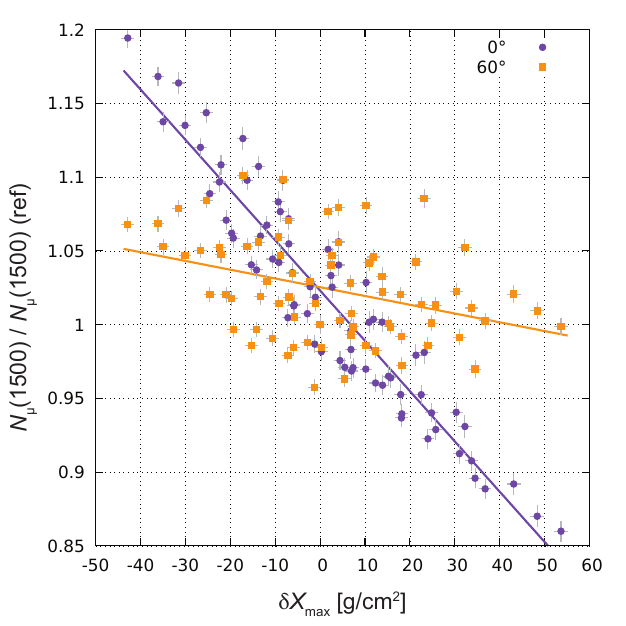}
    \end{subfigure}
        \caption{The impact of various modifications on \Xmax ($\Xmax\rightarrow\Xmax+\delta\Xmax$) and the number of muons at the ground wrt. unmodified value (ref), represented by either the total number of muons (top left) or by the number of muons at a given distance (500, 1000, 1500~m) from the shower axis (rest of the panels) at the two extreme zenith angles. Linear trends have been fitted to the plots to highlight the changes in the slopes with radial distance. Based on \cite{rumunsko}.}
    \label{fig:xmax-smu-r}
\end{figure*}

To illustrate the importance of the 3-dimensional simulations in the modified \conexD, as compared to the 1-dimensional simulations in \conex, we use plots based on those shown in \cite{rumunsko} (Fig.~\ref{fig:xmax-smu-r}), which show the impact of a whole set of combinations of modifications simultaneously on the depth of shower maximum \Xmax and the number of muons. Two sets of points show simulations at the two extreme zenith angles of the simulation library used. The values of $f_{19}$ for each of the three kinds of modifications run between the extremes used in Section \ref{sec:rep}; the configuration of the simulations is identical to that of Section \ref{sec:ground}, with the ground altitude corresponding to that of the Pierre Auger Observatory and the primary particles being protons at 5 EeV. The number of muons is either taken as the total number of muons (which is also accessible in 1-dimensional \conex simulations) or as the number of muons at 500, 1000 or 1500 meters from the shower axis (in the perpendicular plane). It is clear that while at high zenith angles, the results are largely unchanged by the 3-dimensional information, for lower zenith angles, the total number of muons almost entirely misses the strong anti-correlation between the changes in the muon number and \Xmax. The total number of muons in a UHECR shower is a quantity very difficult to measure experimentally, as all existing detectors are sparse arrays and thus most showers are statistically observed at a distance from their shower core. Since the modifications have considerably different effects at different distances, 3-dimensional information is crucial in the comparison with experimental data.

\begin{table*}
  \caption{Test of commutativity of changes in multiplicity and elasticity: ratio of  $N_\mu$ at 1000 meters from the shower cores ($R_\mu$) and difference in the changes in $X_\mathrm{max}$ ($\Delta (\delta X_\mathrm{max})$) between the two options for the order of the changes, for the most extreme modifications for primary protons at 60 degrees of zenith angle. The fourth and sixth columns express the values in units of the statistical uncertainty for the ratio/difference between two sets of 1000 simulated showers.}
  \centering
  \begin{ruledtabular}

  \begin{tabular}{cccccc}
       Elasticity & 
       Multiplicity  &
       $R_\mu$ &
       $R_\mu-1$ [$\sigma$] & 
       $\Delta (\delta X_\mathrm{max})$ [g/cm$^2$] &
       $\Delta (\delta X_\mathrm{max})$ [$\sigma$] 
        \\
    \colrule
    0.6 & 0.6 & 1.004 & 0.4 & 1.4 & 0.6  \\ 
    0.6 & 1.7 & 1.002 & 0.2 & 1.1 & 0.4  \\ 
    1.5 & 0.6 & 1.011 & 1.3 & $-4.0$ & $-1.3$  \\    
    1.5 & 1.7 & 1.006 & 0.7 & 2.7 & 0.9  \\ 
  \end{tabular}
  \end{ruledtabular}

  \label{tab:commutativity}
\end{table*}

\begin{figure*}
    \includegraphics[width=.9\linewidth]{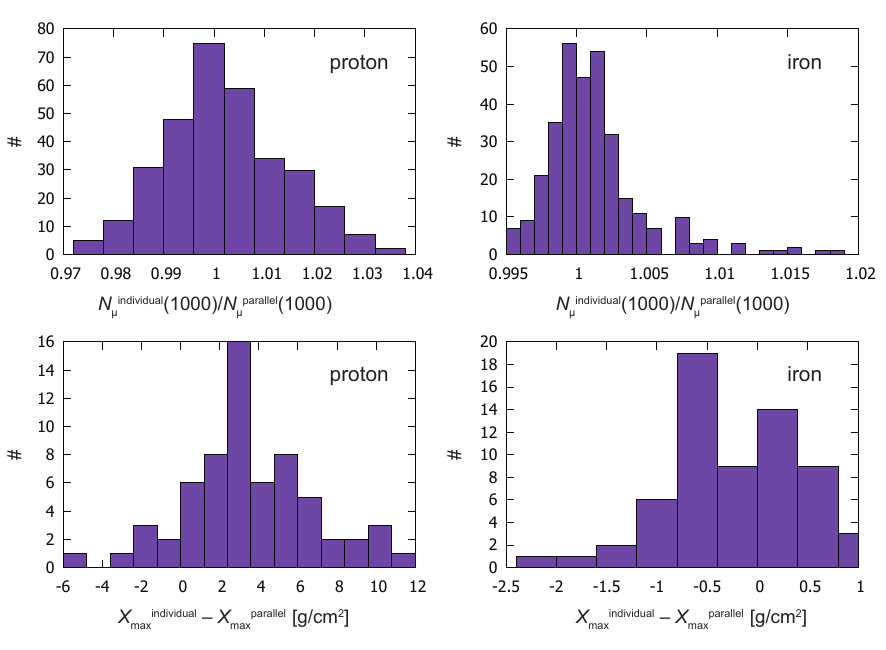}
    \caption{Comparison between the results of parallel modifications and of a linear combination of individual modifications of cross-section, elasticity and multiplicity – the ratio of numbers of muons at 1000 meters from the shower axis (first row) and the difference in the depths of shower maxima (second row), for primary proton (left) and iron nucleus (right). The widths of the bins correspond to the uncertainty of the determination of the given quantity from the library of 1000 showers per bin for $N_\mu$ and 3000 showers for $X_\mathrm{max}$.}
    \label{fig:composition_hist}
\end{figure*}

\section{Impact of parallel modifications}

One of the improvements with respect to the original implementation of \cite{Ulrich:2010rg} is the ability to make modifications of all three parameters -- cross section, elasticity and multiplicity -- in parallel. While such capability is indispensable if individual showers or exact distributions of observables are needed, for mean observables one could, in principle, apply the modifications individually and then add the changes of the observable together. Using the same library of modified simulations as for Fig.~\ref{fig:xmax-smu-r}, we can easily compare the outcomes of the two approaches. For each set of simulations where at least two parameters are modified, we take the corresponding simulations with individual modifications of these parameters and calculate $N_\mu^\mathrm{individual}$ as the product of the individual changes in $N_\mu$, and $X_\mathrm{max}^\mathrm{individual}$ as the sum of the individual $X_\mathrm{max}$ shifts. 

Fig.~\ref{fig:composition_hist} shows the histograms of the ratios/differences of these values with respect to those obtained with the parallel modifications of all parameters. The widths of the bins have been chosen so that they correspond to the statistical uncertainty of the determination of the relevant quantity from the available library, that is, 1000 showers for a specific set of modifications and zenith angle for $N_\mu$ and 3000 showers for the specific set of modifications for $X_\mathrm{max}$ (as $X_\mathrm{max}$  does not depend on the zenith angle). Fig.~\ref{fig:composition_hist} includes the results specifically for muons at 1000 meters, but the histograms are almost identical for all muons or other radial distances. Even if the difference in the predicted number of muons between parallel and individual modifications is small, it is quite significant even for a sample of 1000 showers, in particular for iron nuclei, where the intrinsic fluctuations are very small -- the tail of large differences comes almost exclusively from combining very low multiplicity with very low elasticity. For $X_\mathrm{max}$, the differences are large in particular for primary protons. In general, we conclude that doing the simulations in parallel is worthwhile if precision at the level of a few per cent in $N_\mu$ and less than 10 g/cm$^2$ in $X_\mathrm{max}$ is required.

Since we have established the importance of parallel modifications, we need to address the fact that the resampling of the secondary particles produced in each interaction needed to alter the multiplicity and elasticity is carried out in two separate steps, each for each modified quantity, and thus the result could, in principle, be affected by the order of these steps. This is alleviated by the fact that the leading particle is never included in the multiplicity resampling, but the ``commutativity" of the particular implementation of the  modifications still needs to be tested. To do that, we have simulated 1000 proton-induced showers (at 60 degrees of zenith angle) for each of the four combinations of the extreme elasticity and multiplicity changes from the library used in Fig.~\ref{fig:xmax-smu-r}, with the order of the resampling for elasticity and multiplicity switched and compared the changes in $N_\mu$ at 1000 meters from the shower axis and $X_\mathrm{max}$ with respect to the original order (Table~\ref{tab:commutativity}). The effect of the order of operations is small and compatible with the statistical uncertainties when comparing two sets of 1000 simulated showers (a $1.3\sigma$ difference is expected for almost 60 \% of sets of four events).
 
\FloatBarrier
\section{Conclusions}

We have shown that \conexD simulations are consistent with \conex and \corsika simulations, even when the lower energy threshold of \conex Monte Carlo simulations is set to the very low values needed for maximum flexibility when implementing modified hadronic interactions. The agreement is very good for the mean longitudinal shower profiles and lateral ground distributions of various quantities, and generally satisfactory (within 10 \% or less), except for the fluctuations of high-energy muons ($>5$\,GeV) far from the shower core and EM energy closer to the shower core. We have further shown that the effects of modifications  of cross section, elasticity and multiplicity which we have implemented in \conexD are consistent with simulations using \conex for quantities where such comparisons are applicable. We have presented an example using the changes in the number of muons at different  distances from the shower axis, which shows that 1-dimensional simulations do not provide the full picture when it comes to the effect of modifications on observable quantities. Finally, we have shown that parallel modifications of multiple parameters are important as they lead to different predictions than simple linear combinations of individual effects, albeit only when a high precision in the predictions is required.

The results of this paper show that the implementation of the ad-hoc modifications of hadronic interactions in \conexD is consistent and can be used for the study of the impact of these modifications, if the relevant limitations are respected. These limitations lie mainly in the description of the behavior of muons with energies above 5 GeV at ground and their relative fluctuations, but for specific primary/zenith angle combinations, the EM energy and total charge particle fluctuations can deviate by up to 15 \%. If high precision is required, it is important to note that the muon number at ground is systematically overestimated in \conexD by 3--4 \% across radial distances, primary types and zenith angles. We are currently preparing a publication where the physics implications of modified properties of interactions of ultra-high-energy cosmic-ray showers will be studied in detail and compared with experimental data and other hadronic interaction models.

\FloatBarrier
\section*{Acknowledgements}

This work was supported by the following sources: Czech Science Foundation: 21-02226M, Czech Academy of Sciences: LQ100102401, Regional funds of EU/MEYS: OPJAK FORTE CZ.02. 01.01/00/22\_008/0004632. 

The simulated data used in this work are openly available \cite{ebr_2026_18282786}, including a steering card template for 
\corsika, which however requires a version of \corsika code yet to be made publicly available for execution.

\bibliography{bibliography-mochi.bib}

\end{document}